\documentclass[preprint,showpacs,amsmath,amssymb,superscriptaddress,floatfix,prc]{revtex4}
\usepackage{graphicx}
\usepackage{dcolumn}
\usepackage{bm}
\usepackage{color}

\begin{document}
\newcommand{\dAMg}{d_{\alpha\mbox{-}^{24}{\rm Mg}}}
\newcommand{\dCO}{d_{^{12}{\rm C}\mbox{-}^{16}{\rm O}}}
\newcommand{\Cm}{{{\rm C}_m}}
\newcommand{\Cn}{{{\rm C}_n}}
\newcommand{\CmCn}{{\Cm\mbox{-}\Cn}}
\newcommand{\AMg}{{\alpha\mbox{-}^{24}{\rm Mg}}}
\newcommand{\CO}{{^{12}{\rm C}\mbox{-}^{16}{\rm O}}}
\newcommand{\bfsigma}{{\boldsymbol \sigma}}
\newcommand{\bra}[1]{\langle #1 |}
\newcommand{\ket}[1]{| #1 \rangle}
\newcommand{\ovlp}[2]{\langle #1 | #2 \rangle}
\newcommand{\vect}[1]{\mathbf #1}
\newcommand{\Phirm}{{\rm \Phi}}

\title{Cluster structures and superdeformation in $^{28}$Si}
 \author{Yasutaka Taniguchi}
 \affiliation{RIKEN Nishina Center for Accelerator-Based Science, RIKEN, Wako 351-0198, Japan}
 \author{Yoshiko Kanada-En'yo}
 \affiliation{Yukawa Institute for Theoretical Physics, Kyoto University, Kyoto 606-8502, Japan}
 \author{Masaaki Kimura}
 \affiliation{Creative Research Initiative ``Sousei,'' Hokkaido University, Sapporo 001-0021, Japan}
 \date{\today}
 \begin{abstract}
  We have studied  positive-parity states of $^{28}$Si using antisymmetrized molecular dynamics (AMD) and multi-configuration mixing (MCM) with constrained variation. 
  Applying constraints to the cluster distance and the quadrupole deformation of the variational calculation, we have obtained basis wave functions that have various structures such as $\alpha$-$^{24}$Mg and $^{12}$C-$^{16}$O cluster structures as well as deformed structures. 
  Superposing those basis wave functions, we have obtained a oblate ground state band, a $\beta$ vibration band, a normal-deformed prolate band, and a superdeformed band.  It is  found that the normal-deformed and superdeformed bands contain large amounts of the $^{12}$C-$^{16}$O and $\alpha$-$^{24}$Mg  cluster  components,  respectively. The results also suggest the presence of two excited bands with the developed $\alpha$-$^{24}$Mg cluster structure, where the inter-cluster motion and the $^{24}$Mg-cluster deformation play important roles.
 \end{abstract}
  
 
 \maketitle
  
 \section{Introduction}
Clustering plays critical roles in excited states of  $p$-shell and very light $sd$-shell nuclei  such as $^{16}$O and $^{20}$Ne\cite{PTPS.52.89,PTPS.68.29}.  
 In spite of the importance of clustering in the $A\lesssim 20$ region, its role in the $A\gtrsim 30$ region has not been studied enough. 
 In such a heavier mass region, other effects such as deformations are considered to become more important than in a lighter mass region. 
  Recently, based on AMD calculations, there has been discussion that both mean-field and cluster aspects play important roles in the excited states of $^{32}$S, $^{40}$Ca and $^{44}$Ti\cite{taniguchi:044317,Kimura200658,PhysRevC.69.051304,PhysRevC.72.064322}. 
  
  In order to understand the mean-field and cluster aspects in the heavier system, $^{28}$Si is an important case, because $^{28}$Si has a rich variety of structures in its excited states from view points of both clustering and deformations. 
  In the ground state band and the low-lying excited band, the coexistence of the prolate and the oblate deformation has been studied for quite some time. 
  The ground state band is oblately deformed, while the excited band built on the $0_3^+$ state at 6.69 MeV is considered to have prolate deformation (prolate ND)\cite{GSB+81,PhysRevLett.46.1559,Sheline1982263,DasGupta1967602}. 
  Among these states, the $\beta$ vibration of the oblate deformed ground state band generates the band built on the $0_2^+$ state at 4.98 MeV\cite{Sheline1982263,DasGupta1967602}. 
  Furthermore, Kubono {\it et al.} have proposed a largely deformed band called the ``excited prolate'' band based on a $^{12}$C($^{20}$Ne,~$\alpha$) reaction\cite{PhysRevC.33.1524,Kubono1986461,Kubono1981320}. 
  This band assignment, however, is not confirmed yet, because intra-band electromagnetic transitions have not been observed. 

  In the highly excited states of $^{28}$Si, the cluster aspects have been discussed. 
  The excited states from $E_x=18$ to 30 MeV, which have been observed by $^{24}$Mg($^6$Li,~$d$), $^{24}$Mg($\alpha$,~$\alpha$) and $^{24}$Mg($\alpha$,~$\gamma$) reactions are suggested as candidates for the rotational band members that have an $\alpha$-$^{24}$Mg cluster structure\cite{AGG+90,Cseh198243}, 
  though a detailed theoretical study of $\alpha$-$^{24}$Mg cluster states has not been done. 
  Another cluster feature of $^{28}$Si is $^{12}$C-$^{16}$O clustering, which has been intensively investigated experimentally and theoretically. 
  Around the excitation energy region of 30--50 MeV,  the $^{12}$C~+~$^{16}$O molecular resonances 
  have been experimentally observed by elastic, inelastic, other exit channels and fusion cross sections\cite{James1976177,Charles1976289,Froehlich1976408w,Baye1977176,0954-3899-29-4-307,PhysRevC.63.034311,PhysRevC.63.034315}. 
  On the theoretical side, the $^{12}$C-$^{16}$O molecular resonances and their relation to the low-lying prolate deformed states have been studied by microscopic and macroscopic cluster models\cite{Baye1976445,PTP.74.1053,Ohkubo2004304}.  
  By the $^{12}$C-$^{16}$O potential model, the prolate ND band and observed $^{12}$C-$^{16}$O molecular resonances are reproduced\cite{Ohkubo2004304}. 
  
  We aim in the present work to investigate the nature of excited states of $^{28}$Si, focusing on clustering and deformations in a unified manner. 
  Shape coexistence and $\beta$-vibration are studied, and those states are discussed in relation to cluster components of $\alpha$-$^{24}$Mg and $^{12}$C-$^{16}$O clustering. 
  We also discuss the possible existence of the superdeformation and developed $\alpha$-$^{24}$Mg cluster states. 
  
  In this study, we apply a theoretical framework of deformed-basis antisymmetrized molecular dynamics (deformed-basis AMD) + multi-configuration mixing (MCM) with constrained energy variation. 
  The deformed-basis AMD wave function enables us to describe both clustering and deformation phenomena in a unified manner\cite{taniguchi:044317,PhysRevC.69.044319,PTP.112.475}. 
  In order to study excited states, we first perform 
  energy variation under two kinds of constraints to obtain basis functions. 
  We shall call these constraints  $\beta$ and $d$ constraints that are imposed on the quadrupole deformation $\beta$ and distance $d$ between the centers of mass of clusters, respectively. Then we carry out the MCM by superposing the basis wave functions to obtain energy levels and wave functions of the ground and excited states. The method of deformed-basis AMD + MCM with the $\beta$ and $d$ constraints has been applied already to $^{40}$Ca and proved to be efficient for describing various cluster states and deformed states as well as clustering correlation in the deformed states.\cite{taniguchi:044317}
  In the present study of $^{28}$Si, 
  we naturally adopt the $\alpha$-$^{24}$Mg and $^{12}$C-$^{16}$O clustering for the $d$ constraint, because the $\alpha$-$^{24}$Mg and $^{12}$C-$^{16}$O clustering correlations are expected to play important roles as mentioned before.
  The obtained wave functions are analyzed in order to investigate the nature of the states. 
  
  The present paper is organized as follows: 
  In Sec.~\ref{sec:framework}, the framework of deformed-basis AMD + MCM is explained briefly. 
  Results of energy variation imposing two kinds of constraints and analysis of the obtained wave functions are presented in Sec.~\ref{sec:energy_variation}. 
  In Sec.~\ref{sec:band_structures}, results of MCM and structures  of low-lying states are discussed. 
  Clustering aspects are discussed in Sec.~\ref{sec:cluster_correlations}. 
  Finally, we give a summary and conclusions in Sec.~\ref{sec:summary}. 

  \section{Framework}
  We have used the theoretical framework of deformed-basis AMD + MCM with constraints.\cite{PhysRevC.69.044319} 
  The details are presented in Refs.~\onlinecite{taniguchi:044317}, \onlinecite{PTP.112.475}, and \onlinecite{PhysRevC.56.1844}. 
  \label{sec:framework}
  \subsection{Wave Function and Hamiltonian}
  The deformed-basis AMD
  wave function is a Slater determinant of triaxially deformed Gaussian wave packets, 
  \begin{subequations}
    \label{AMD}
    \begin{eqnarray}
      \ket{\Phi_{\rm int}}& = &\hat{\cal A} \ket{\varphi_1,\  \varphi_2,\cdots,\varphi_A}, \\
\ket{\varphi_i}& = &\ket{\phi_i,\  \chi_i,\   \tau_i}, \\
\ovlp{\vect{r}}{\phi_i}& = &\prod_{\sigma = x, y, z} \left( \frac{2\nu_\sigma}{\pi} \right)^{\frac{1}{4}} \exp \left[ - \nu_\sigma \left( r_\sigma - \frac{Z_{i\sigma}}{\sqrt{\nu_\sigma}} \right)^2 \right], \nonumber\\
\\
\ket{\chi_i}& = &\alpha_i \ket{\uparrow} + \beta_i \ket{\downarrow},\\
\ket{\tau_i}&  = &\ket{p}\  {\rm or}\  \ket{n}. 
    \end{eqnarray}
  \end{subequations}
  Here, the complex parameters $\vect{Z}_i$, which represent the centroids
  of the Gaussian wave packets in phase space, take independent values for each single-particle wave function.  The width parameters $\nu_x$, $\nu_y$, and
  $\nu_z$ are real parameters and take independent values for each of the
  $x$-, $y$- and $z$-directions, but are common for all nucleons.  The
  spin part $\ket{\chi_i}$ is parameterized by $\alpha_i$ and $\beta_i$ and the
  isospin part $\ket{\tau_i}$ is fixed as $\ket{p}$ (proton) or $\ket{n}$ (neutron).  The
  values  $\{\vect{Z}_i$, $\alpha_i$, $\beta_i\}$ $(i=1,\cdots,A)$, $\nu_x$, $\nu_y$ and $\nu_z$ are
  variational parameters and are optimized by energy variation as
  explained  below. 
  
  The trial wave function in the energy variation with constraints is
  a parity-projected wave function,  
  \begin{equation}
    \ket{\Phi^\pi} = \frac{1+\pi \hat{P}_r}{2} \ket{\Phi_{\rm int}}, 
\end{equation}
  where $\pi$ is parity 
  and $\hat{P}_r$ is the parity operator.  In this study, we will
  discuss positive-parity states.  
  
  The Hamiltonian is,
  \begin{equation}
    \hat{H} = \hat{K} + \hat{V}_{\rm N} + \hat{V}_{\rm C} - \hat{K}_{\rm G}, 
  \end{equation}
  where $\hat{K}$ and $\hat{K}_{\rm G}$ are the kinetic energy and the
  energy of the center of mass motion, respectively, and $\hat{V}_{\rm N}$
  is the effective nucleon-nucleon interaction.  We have used the Gogny D1S force\cite{Berger1991365}, which is one of the widely used effective forces for (beyond-)mean-field approaches. 
  It is consists of the finite-range and zero-range density dependent two-body central terms and the zero-range two-body spin-orbit term. 
  The form of the Gogny D1S force is given as, 
\begin{widetext}
 \begin{eqnarray}
  \hat{V}_\mathrm{N} &=& \sum_{i < j} \hat{v}^\mathrm{N}_{ij}, \\
  \hat{v}^\mathrm{N}_{12} &=& \sum_{n = 1}^2 e^{- (\hat{\vect{r}}_1 - \hat{\vect{r}}_2)^2 / \mu_n^2} (W_n + B_n \hat{P}^\sigma - H_n \hat{P}^\tau - M_n \hat{P}^\sigma \hat{P}^\tau) \nonumber \\
  && + i W_0 (\hat{\bfsigma}_1 + \hat{\bfsigma}_2) \hat{\vect{k}} \times \delta(\hat{\vect{r}}_1 + \hat{\vect{r}}_2) \hat{\vect{k}} 
   + t_3 (1 + \hat{P}^\sigma) \delta(\hat{\vect{r}}_1 - \hat{\vect{r}}_2) \rho^{1/3} \left(\frac{\hat{\vect{r}}_1 + \hat{\vect{r}}_2}{2}\right).\label{GognyD1S} 
 \end{eqnarray}  
\end{widetext}  
Where the $\hat{P}^\sigma$ and $\hat{P}^\tau$ are exchange operators of spin and isospin parts, respectively, the $\hat{\bm{\sigma}}$ is the spin operator, and the $\hat{\vect{k}}$ is the operator of the relative momentum $\hat{\vect{k}} = (\hat{\vect{p}}_1 - \hat{\vect{p}}_2) / 2 \hbar$. 
Force parameters of the Gogny D1S force are
$\mu_1 = 0.7\ \mathrm{fm}$, $W_1 = -1720.30\ \mathrm{MeV}$, $B_1 = 1300.00\ \mathrm{MeV}$, $H_1 = -1813.53\ \mathrm{MeV}$, $M_1 = 1397.60\ \mathrm{MeV}$,  $\mu_2 = 1.2\ \mathrm{fm}$, $W_2 = 103.64\ \mathrm{MeV}$, $B_2 = -163.48\ \mathrm{MeV}$, $H_2 = 162.81\ \mathrm{MeV}$, $M_2 = -223.93\ \mathrm{MeV}$, $W_0 = -130\ \mathrm{MeV\ fm^5}$ and $t_3 = 1390.60\ \mathrm{MeV\ fm^4}$. 
This force has been proved to reproduce the binding energy in the wide mass region systematically. 
The Coulomb force $\hat{V}_{\rm C}$ is approximated by a sum of seven Gaussian functions.  
  
  \subsection{Energy Variation}
  We have performed energy variation and optimized the variational parameters
  included in the trial wave function [Eqs. (\ref{AMD})] to find the state
  that minimizes the energy of the system $E^\pi$,  
  \begin{equation}
    E^\pi = \frac{\bra{\Phi^\pi}\hat{H}\ket{\Phi^\pi}}{\ovlp{\Phi^\pi}{\Phi^\pi}} + V_{\rm cnst}. 
  \end{equation}
  Here, we add the constraint potential $V_{\rm cnst}$ to the expectation
  value of Hamiltonian $\hat{H}$ in order to obtain  energy-minimum
  states under the optional constraint condition.  In this study, we
  employ two types of constraints, one on the  quadrupole
  deformation parameter $\beta$ ($\beta$ constraint) and other on the distance
  between clusters' centers of mass $d$ ($d$ constraint) by using 
  the potential $V_{\rm cnst}$,  
  \begin{equation}
    V_{\rm cnst} = 
    \left\{
    \begin{array}{ll}
      v_{\rm cnst}^\beta ( \beta - \tilde{\beta} )^2 & \mbox{for $\beta$ constraint,}\\
      v_{\rm cnst}^d ( d_{{\rm C}_m\mbox{-}{\rm C}_n} - \tilde{d}_{{\rm C}_m\mbox{-}{\rm C}_n})^2          & \mbox{for $d$ constraint.}
    \end{array}
    \right. . 
  \end{equation}
  Here $\beta$ is the matter quadrupole deformation parameter, and $d_{{\rm C}_m\mbox{-}{\rm C}_n}$
  is the distance between the clusters' centers of mass ${\rm C}_m$ and
  ${\rm C}_n$,  
  \begin{eqnarray}
    &d_\CmCn = \left| \vect{R}_\CmCn \right|,&\\
    &\vect{R}_\CmCn = \vect{R}_\Cm - \vect{R}_\Cn ,&\\ 
    &R_{\Cn\sigma} = \frac{1}{A_\Cn} \sum_{i\in {\rm C}_n} \frac{{\rm Re} Z_{i\sigma}}{\sqrt{\nu_\sigma}},&  
  \end{eqnarray}
  where $A_\Cn$ is the mass number of cluster ${\rm C}_n$ and the
  expression $i\in {\rm C}_n$ means that the $i$th nucleon is contained in
  cluster C$_n$.  It should be noted that the $\sigma\ (=x, y, z)$
  component of the spatial center of the single-particle wave function
  $\ket{\varphi_i}$ is $\frac{{\rm Re} Z_{i\sigma}}{\sqrt{\nu_\sigma}}$. 
  When sufficiently large values are chosen for $v_{\rm cnst}^\beta$ and $v_{\rm
    cnst}^d$, the resultant values $\beta$ and $d_{{\rm C}_m\mbox{-}{\rm
      C}_n}$ of energy variation become $\tilde{\beta}$ and $\tilde{d}_\CmCn$, respectively.  We constrain the $\dAMg$ and $\dCO$
  values for the $d$ constraint.  In each calculation of energy variation, we
  constrain one of the values $\beta$, $\dAMg$ and $\dCO$. 
  Other quantities such as triaxiality $\gamma$ are not constrained and optimized by the energy variation. 
  
  The energy variation with the AMD wave function is carried out using the
  frictional cooling method.\cite{PTP.93.115} 
  The time evolution equation for the complex parameters $\vect{Z}_i, \alpha_i$ and $\beta_i$ is  
  \begin{equation}
    \frac{dX_i}{dt} = - \mu_X \frac{\partial E^\pi}{\partial X_i^*},\  (i=1, 2, \cdots, A), 
  \end{equation}
  where $X_i$ is $\vect{Z}_i, \alpha_i$ or $\beta_i$, and the time evolution equation for the real parameters $\nu_x, \nu_y$, and $\nu_z$ is  
  \begin{equation}
    \frac{d\nu_\sigma}{dt} = - \mu_\nu \frac{\partial E^\pi}{\partial
      \nu_\sigma},\  (\sigma = x, y, z).  
  \end{equation}
  The quantities $\mu_X$ and $\mu_\nu$ are arbitrary positive real numbers. 
  The energy of the system decreases as time progresses, and after a
  sufficient number of time steps, we obtain a  minimum energy state under the condition satisfying the given constraint.

  \subsection{Angular Momentum Projection and Multi-configuration Mixing} 
  After performing the constraint energy variation for $\ket{\Phi^\pi}$, we
  superpose the optimized wave functions by employing the quadrupole
  deformation parameter $\beta$ and the distances between the centers of mass
  among clusters $d_{{\rm C}_m\mbox{-}{\rm C}_n}$ for the ${\rm C}_m\mbox{-}{\rm C}_n$
  configurations, 
  \begin{eqnarray}
    \ket{\Phi^{J^\pi}_M} 
    = 
    \sum_K &\hat{P}_{MK}^{J^\pi}&
    \left(
    \sum_i f_{iK}^\beta \ket{\Phi^\beta_i} 
    \right. 
    \nonumber\\
    &&
    \left.
    + \sum_{i,{{\rm C}_m\mbox{-}{\rm C}_n}} f_{iK}^{d_{{\rm C}_m\mbox{-}{\rm C}_n}}
    \ket{\Phi^{d_{{\rm C}_m\mbox{-}{\rm C}_n}}_i} 
    \right) \label{eq:HW}
  \end{eqnarray}
  where $\hat{P}_{MK}^{J^\pi}$ is the parity and total angular momentum
  projection operator, and $\ket{\Phi^\beta_i}$ and
  $\ket{\Phi^{d_{{\rm C}_m\mbox{-}{\rm C}_n}}_i}$ are optimized wave functions
  with $\beta$  and $d_{{\rm C}_m\mbox{-}{\rm C}_n}$ constraints for the constrained
  values $\tilde{\beta}^{(i)}$ and $\tilde{d}_{{\rm C}_m\mbox{-}{\rm C}_n}^{(i)}$
  respectively.  
  The integrals over the three Euler angles in the total angular momentum projection operator $\hat{P}_{MK}^J$ are evaluated by numerical integration.  
  The mesh widths in numerical integration are $2\pi / 9, \pi / 257$ and $2\pi / 9$ for $\alpha, \beta$ and $\gamma$, respectively. 
  Here the body-fixed $x$-, $y$- and $z$-axis are chosen as $\langle x^2 \rangle \leq \langle y^2 \rangle \leq \langle z^2 \rangle$ for $\gamma < 30^\circ$ wave functions and  $\langle x^2 \rangle \geq \langle y^2 \rangle \geq \langle z^2 \rangle$ for $\gamma > 30^\circ$ ones in the case of $\beta$ constrained wave functions. 
  In the case of $d$-constrained wave functions, the $z$-axis is chosen as the vector $\vect{R}_\CmCn$, which connects the ${\rm C}_m$ and $\Cn$ clusters. 
  The coefficients $f_{iK}^\beta$ and $f_{iK}^{d_{{\rm C}_m\mbox{-}{\rm C}_n}}$ are
  determined by the Hill-Wheeler equation,  
  \begin{equation}
    \delta \left( \bra{\Phi^{J^\pi}_M} \hat{H} \ket{\Phi^{J^\pi}_M} - \epsilon \ovlp{\Phi^{J^\pi}_M}{\Phi^{J^\pi}_M}\right) = 0. 
  \end{equation}
  Then we get the energy spectra and the corresponding wave functions that expressed by the superposition of the optimum wave functions, $\{ \ket{\Phirm^\beta_i} \}$, $\{ \ket{\Phirm^{\dAMg}_i} \}$, and $\{ \ket{\Phirm^{\dCO}_i} \}$. 
  
  \subsection{Single-Particle Orbit and Squared Overlap}
In this subsection, we give the definitions of single-particle orbits
and the squared overlap. 
 These values are useful in analysis of the calculated wave functions.\label{subsec:spooverlap}

  \subsubsection{Single-Particle orbits}
  In order to analyze a AMD wave function $\ket{\Phi_\mathrm{int}}$ from the mean-field description, we have calculated Hartree-Fock-type single-particle orbits $\ket{\tilde{\varphi}_i}$\cite{PhysRevC.56.1844} given by superposition of single-particle wave functions $\ket{\varphi_i}$ of the AMD wave function as follows. 
  
  First, orthonormalized wave functions $\ket{\varphi'_i}$ are obtained by the linear combination of $\ket{\varphi_i}$. 
  Next, the $\ket{\varphi'_i}$ are transformed to $\ket{\tilde{\varphi}_i}$ by the unitary transformation to diagonalize the mean-field Hamiltonian matrix
 \begin{widetext}
  \begin{equation}
   h_{ij} = \bra{\varphi'_i} \hat{t} \ket{\varphi'_j} + \sum_k\bra{\varphi'_i \varphi'_k} (\hat{v}^\mathrm{N} + \hat{v}^\mathrm{C}) (\ket{\varphi'_j \varphi'_k} - \ket{\varphi'_k \varphi'_j}). 
  \end{equation}
 \end{widetext}  
 where $\hat{t}$ is the 1-body operator of the kinetic energy, and the $\hat{v}^\mathrm{N}$ and the $\hat{v}^\mathrm{C}$ are the 2-body operators of the nuclear effective interaction $\hat{v}^\mathrm{N}_{12}$ in Eq.~(\ref{GognyD1S}) and that of the Coulomb force, respectively. 
 Thus obtained $\ket{\tilde{\varphi}_i}$ satisfy the following equations, 
 \begin{widetext}
  \begin{eqnarray}
   & \ovlp{\tilde{\varphi}_i}{\tilde{\varphi}_j} = \delta_{ij},& \\
   & \bra{\tilde{\varphi}_i} \hat{t} \ket{\tilde{\varphi}_j} + \sum_k\bra{\tilde{\varphi}_i \tilde{\varphi}_k} (\hat{v}^\mathrm{N} + \hat{v}^\mathrm{C}) (\ket{\tilde{\varphi}_j \tilde{\varphi}_k} - \ket{\tilde{\varphi}_k \tilde{\varphi}_j}) = e_i \delta_{ij}.& 
  \end{eqnarray}
 \end{widetext}  

  \subsubsection{Squared overlap}
  In order to analyze the contributions to an MCM wave function $\ket{\Phirm^{J^\pi}_M}$ of a subset $X = \{ \ket{\Phirm_i^{(X)}} \}\ (i = 1, 2, \cdots) $  of the total set of basis wave functions, squared overlap $S_X$ is calculated as 
  \begin{equation}
    S_X = \sum_i |\ovlp{\Phirm^{J^\pi}_M}{\tilde{\Phirm}_i^{(X)}}|^2, 
  \end{equation}
  where $\ket{\tilde{\Phirm}_i^{(X)}}$ is an orthonormalized set obtained by the linear combination of $\ket{\Phirm_i^{(X)}}$. 
    
  \section{Structures obtained by constrained energy variation}
  \label{sec:energy_variation}
  
  By energy variation under the constraints on quadrupole deformation parameter $\beta$ and inter-cluster distance $\dAMg$ and $\dCO$ for $\alpha$-$^{24}$Mg and   $^{12}$C-$^{16}$O cluster structures, respectively, energy curves as functions of $\beta$, $\dAMg$ and $\dCO$ are obtained. 
  On the curves, various structures appear. 
  
  \begin{figure}[tbp]
    \begin{center}
      \includegraphics[width=0.5\textwidth]{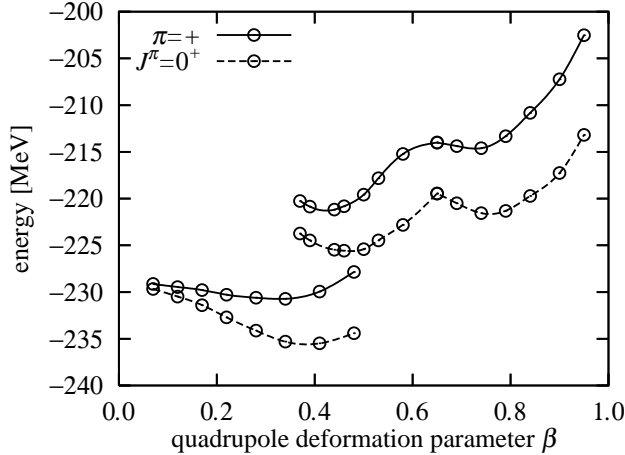}
    \end{center}
    \caption{
      $\beta$-energy curves projected to positive-parity (solid lines) and $J^\pi = 0^+$ states (dashed lines). 
      In smaller and larger $\beta$ regions, oblate and prolate shapes are obtained, respectively. (See text)
    }
    \label{fig:surface_beta}
  \end{figure}

  \begin{figure}[tbp]
   \begin{center}
    \begin{tabular}{ll}
     \includegraphics[width=0.25\textwidth]{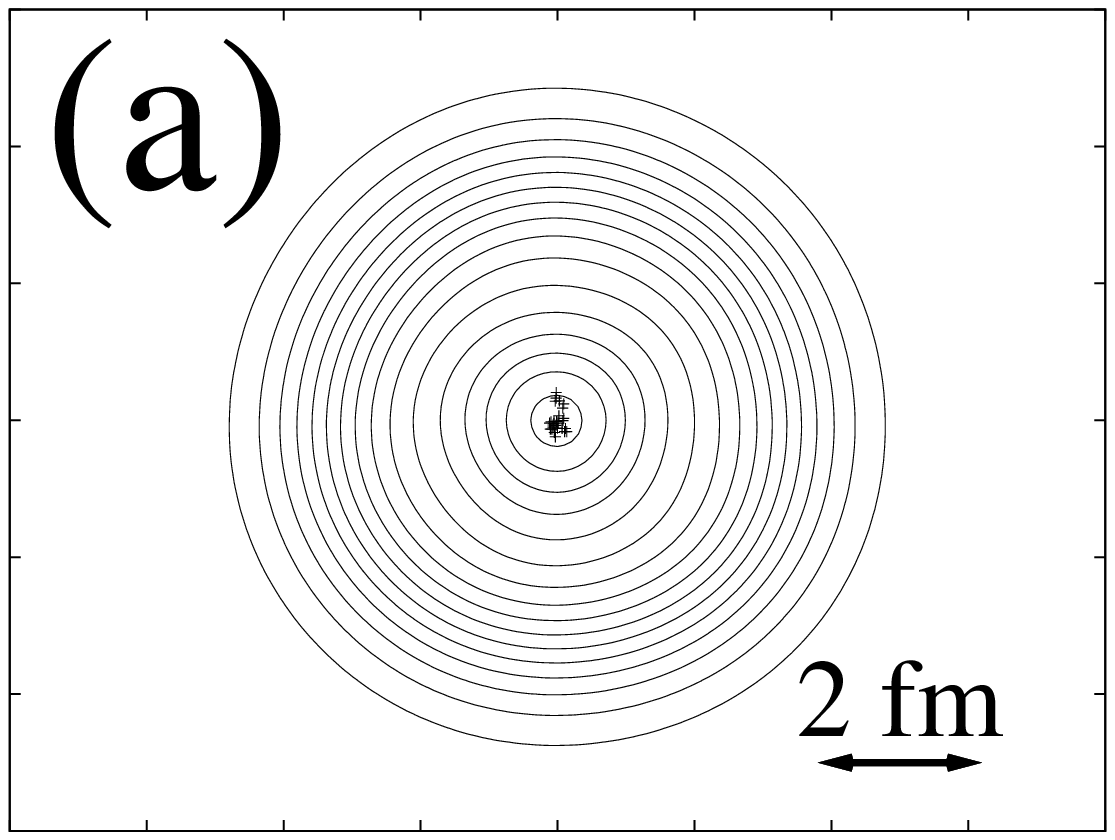}& \\
     \includegraphics[width=0.25\textwidth]{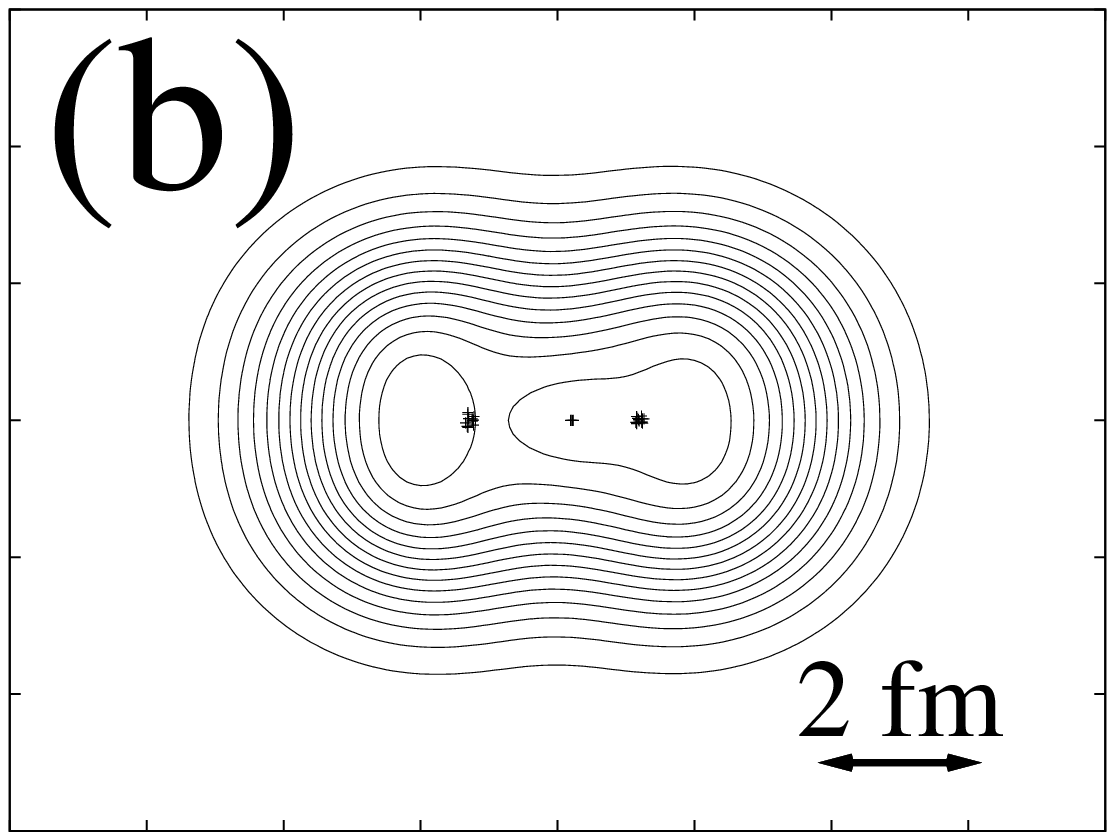} &
	 \includegraphics[width=0.25\textwidth]{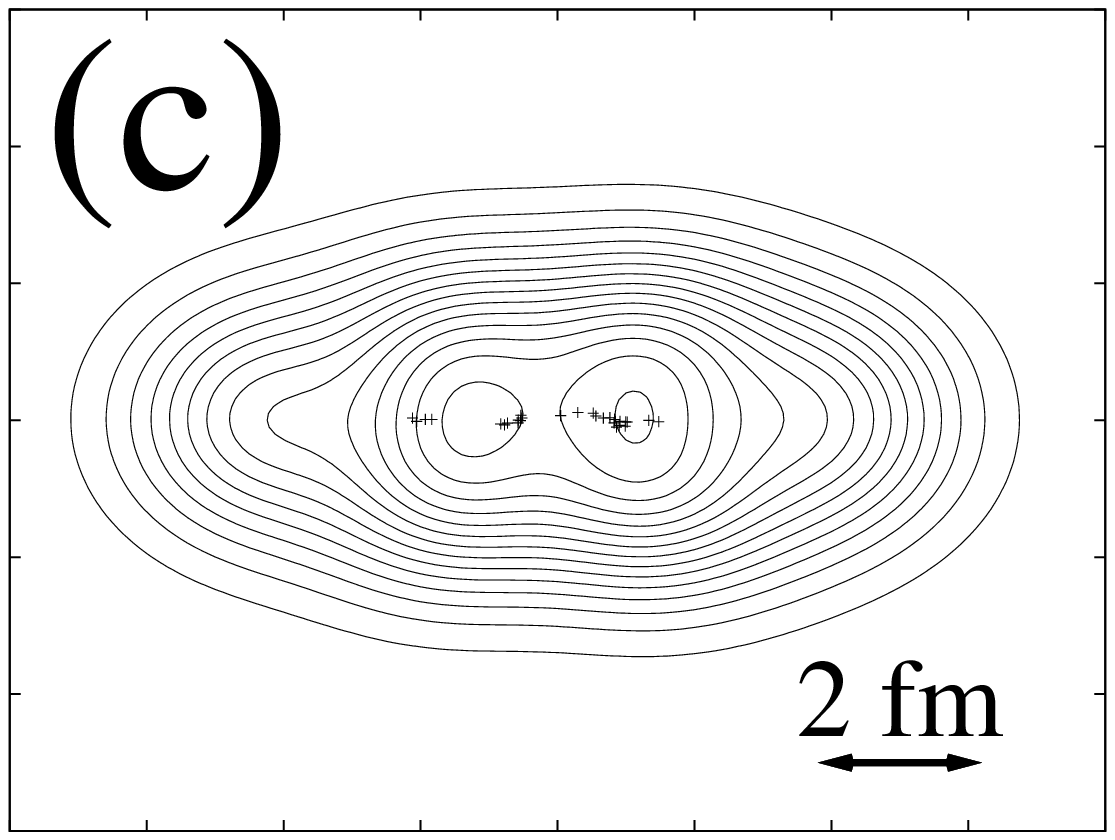} \\
     \includegraphics[width=0.25\textwidth]{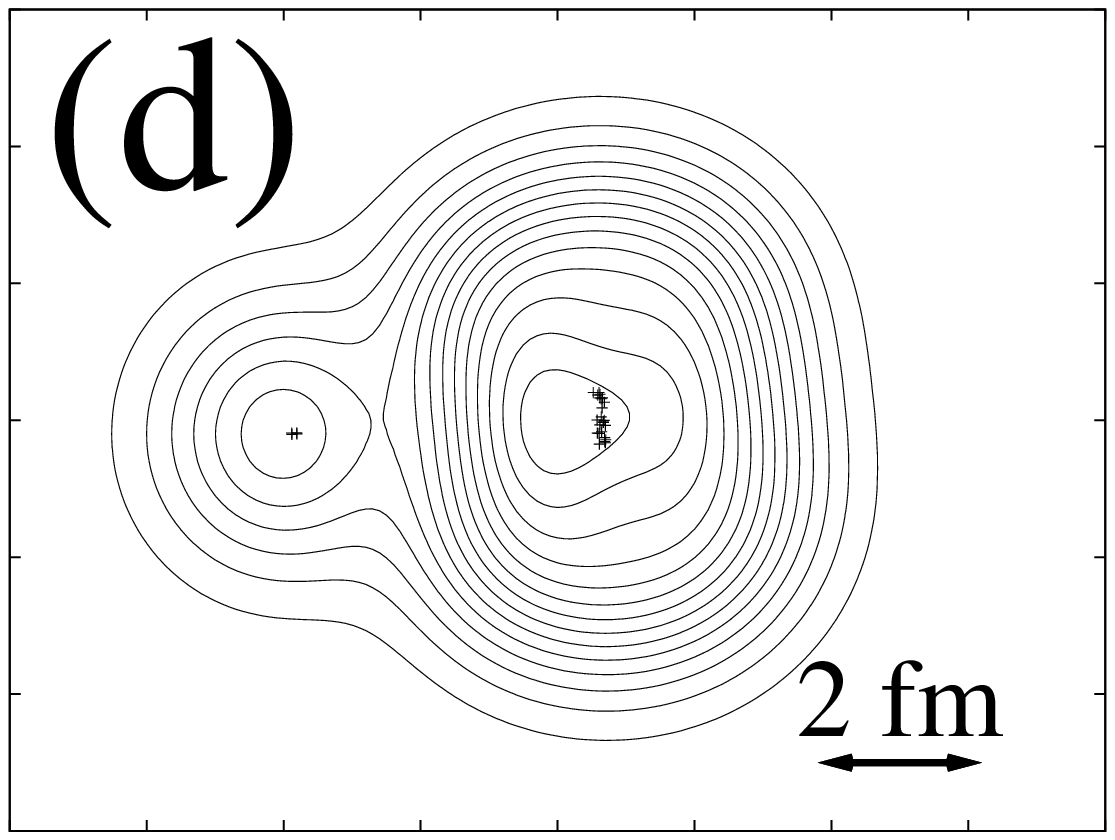} &
	 \includegraphics[width=0.25\textwidth]{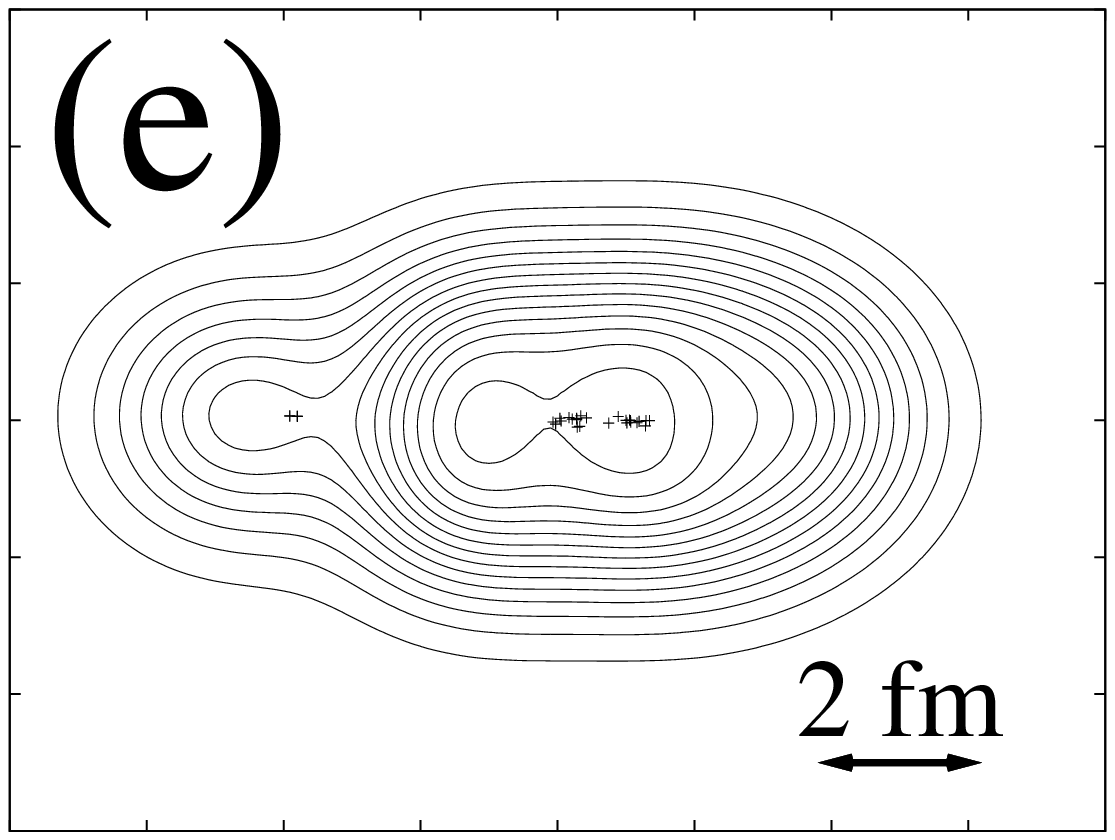} \\
     \includegraphics[width=0.25\textwidth]{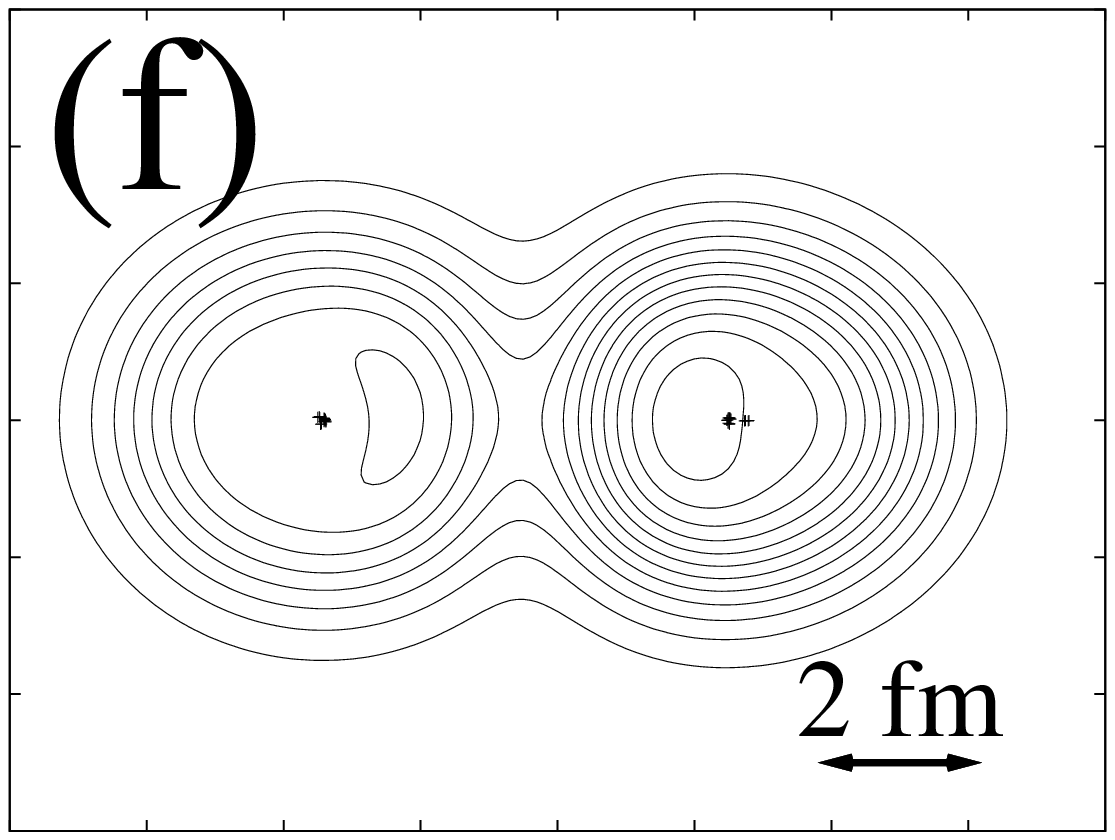} &
	 \includegraphics[width=0.25\textwidth]{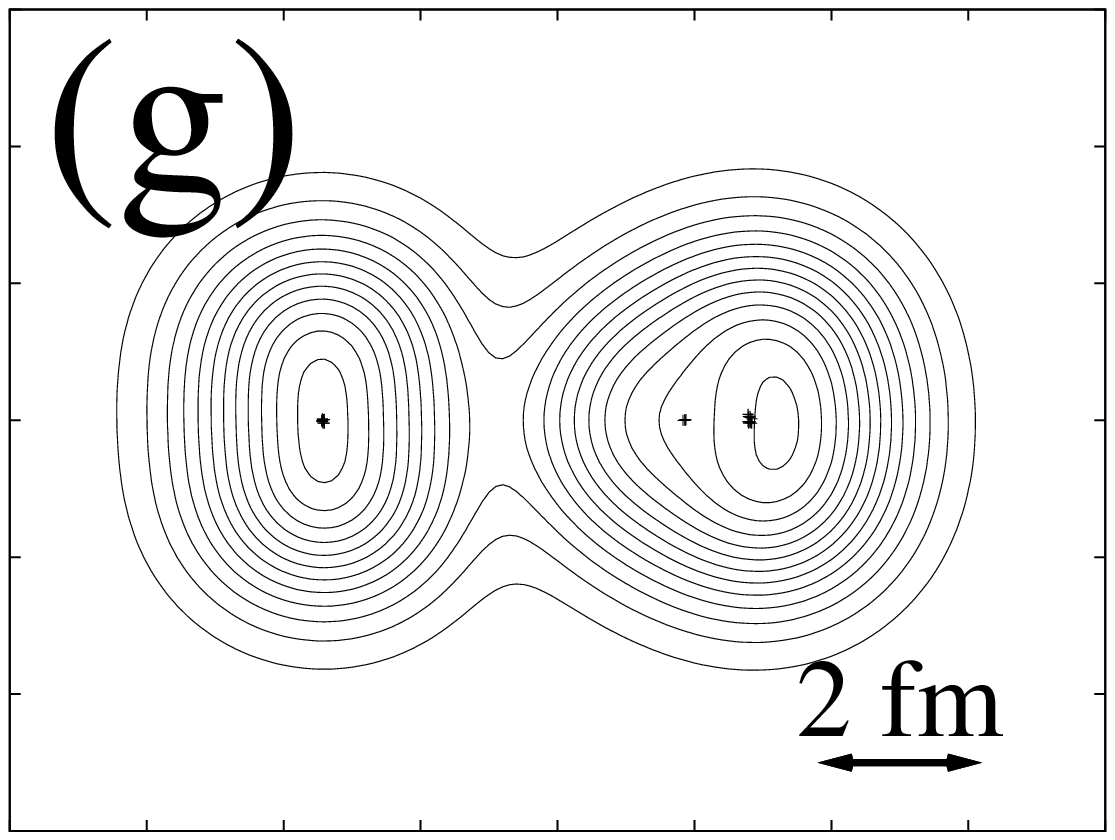} \\
    \end{tabular}
   \end{center}
   \caption{
   Density distribution are shown around (a) oblate, (b) ND ($\beta = 0.46$) and (c) SD local minima ($\beta= 0.79$) for $\beta$-constrained wave functions, (d) type-T and (e) type-A wave functions for $\dAMg$-constrained wave functions ($\dAMg = 4.5$ fm), and (f) type-T and (g) type-A wave functions for $\dCO$-constrained wave functions ($\dCO = 6.0$ fm). 
   Density distributions are projected onto the $yz$-plane, where the $z$- and $x$-axis are major and minor axes, respectively. 
   Symbols ``+'' indicate centroids of wave packets. 
   }
   \label{fig:density}
  \end{figure}

  Figure~\ref{fig:surface_beta} shows the $\beta$-energy curves for the positive-parity states before and after the angular momentum projection to $J^\pi = 0^+$ states. 
  The obtained wave functions always have axially symmetric shapes, though the mass quadrupole deformation parameter $\gamma$ for triaxiality is not constrained and optimized by the energy variation. 
  In the small deformed region $\beta \lesssim 0.5$, the system is oblately deformed and the surface has a shallow minimum. 
  In the large deformed region $\beta \gtrsim 0.4$, the system has prolate deformation and two local minima around $\beta = 0.5$ and 0.7 that we call ND and SD minima in the following discussion. 
  The Skyrme SLy7 force gives similar results\cite{PTP.121.533}. 
  Figures~\ref{fig:density}(a), (b), and (c) show density distributions of the oblate, ND and SD minima obtained with $\beta$ constraint, which have no remarkable cluster structures. 

  \begin{figure}[tbp]
    \begin{center}
      \includegraphics[width=0.5\textwidth]{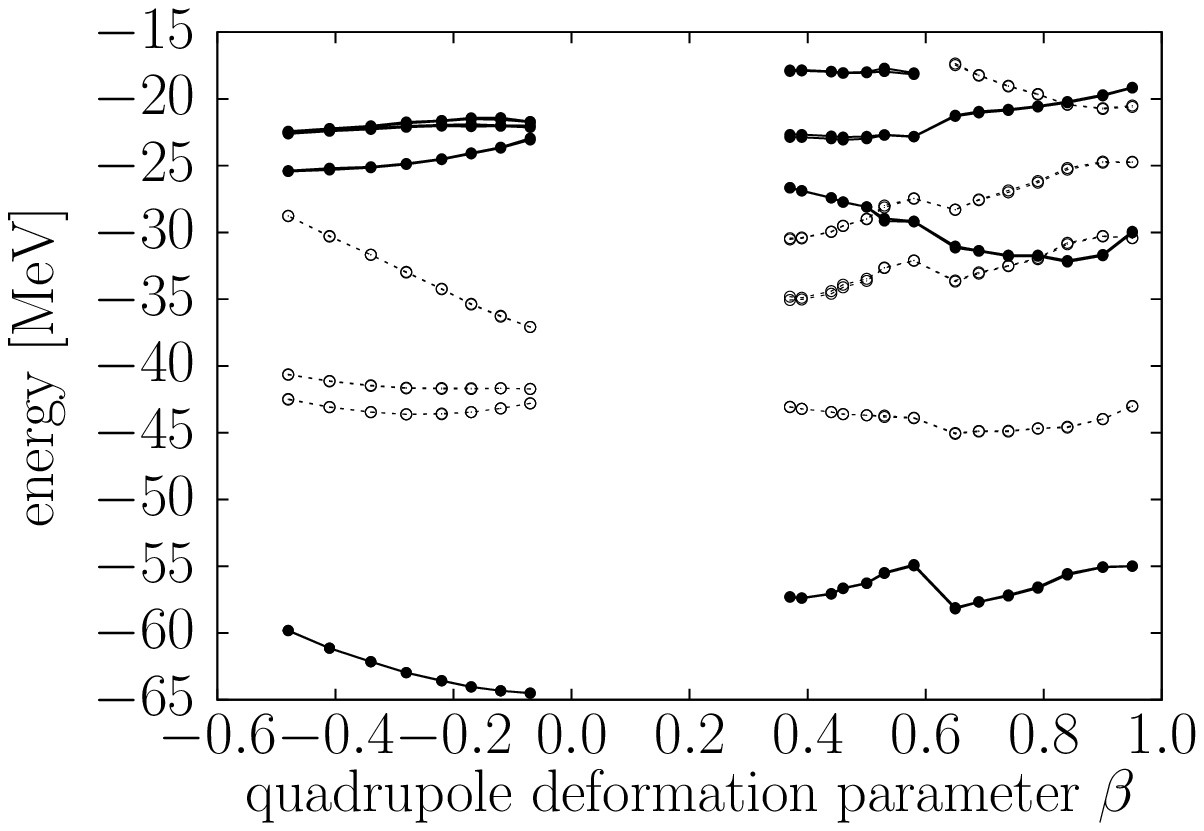}
      \caption{
	Energies of neutrons' single-particle orbits as functions of quadrupole deformation parameter $\beta$. 
	Solid and dashed lines show positive- and negative parity states, respectively. 
      	The $\beta$ values for oblate shapes are defined as negative values. 
      }
      \label{fig:spo}
    \end{center}
  \end{figure}

    \begin{figure}[tbp]
    \begin{center}
     \begin{tabular}{cc}
      \includegraphics[width=0.225\textwidth]{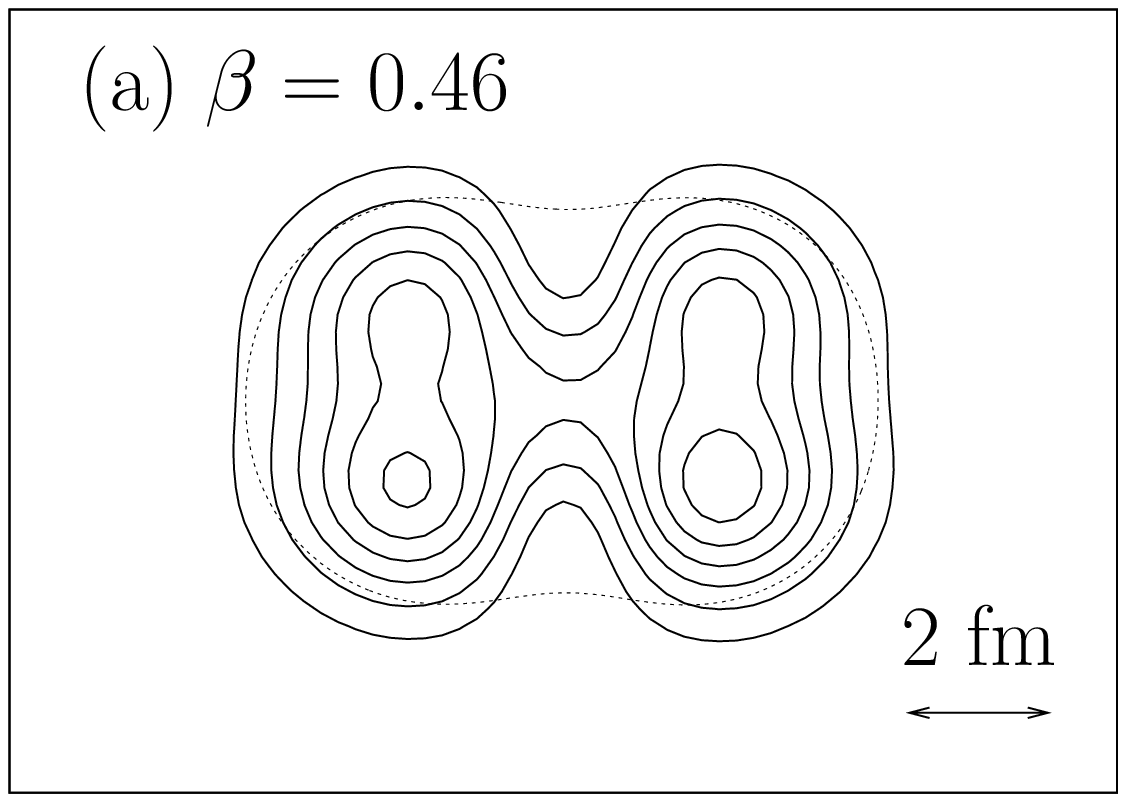}&
      \includegraphics[width=0.225\textwidth]{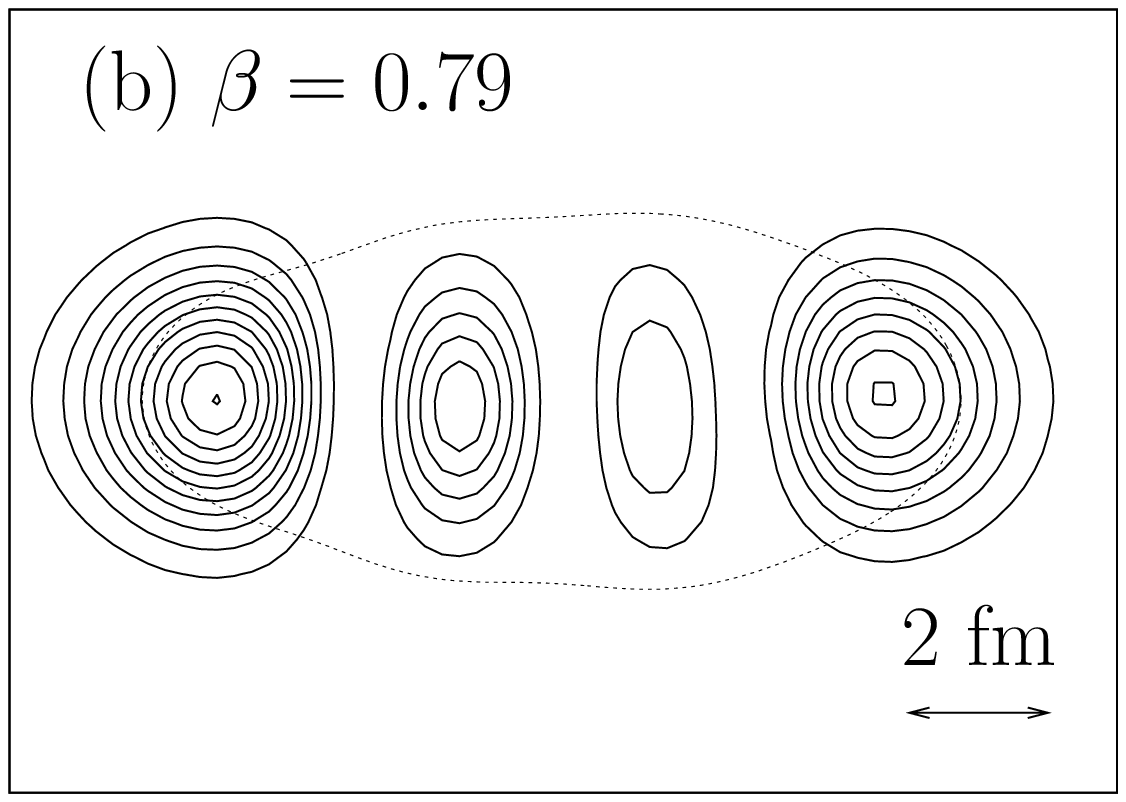}
     \end{tabular}      
\caption{
	Density distributions of the highest-energy single-particle orbits in (a)~$\beta = 0.46$ and (b)~$\beta = 0.79$ wave functions are shown. 
	Solid and dashed lines show density distributions of single-particle orbits and total wave functions, respectively. 
      }
      \label{fig:spo_density}
    \end{center}
  \end{figure}

  In order to study the change of the intrinsic structures as a functions of $\beta$, single-particle orbits are investigated. 
  The single-particle energies $e_i$ for the neutron orbits are shown in Fig.~\ref{fig:spo}. 
  The orbits for protons are similar to those for neutrons qualitatively, but shift by approximately 5 MeV due to the Coulomb energies. 
  Two single-particle orbits are always degenerate due to the time reversal symmetry, and they are approximately the eigenstates parity except for the transitional region from ND to SD minima around $\beta = 0.58$. 
  Positive- and negative-parity orbits are represented by solid and dotted lines, respectively, in Fig.~\ref{fig:spo}. 
  The parity of each single-particle orbits shows that oblate and ND states has $(sd)^{12}$ configurations. 
  In the SD region ($\beta\sim 0.8$), negative-parity orbits intrude, and the system has the $4p4h$ configuration in which four nucleons are promoted into the $pf$-shell across an $N = 20$ shell gap. 
  The promotion of nucleons into the $pf$-shell is confirmed by the density distribution of the highest occupied single-particle orbitals at ND and SD minima (Fig.~\ref{fig:spo_density}). 
  The density distribution at the SD minimum [Fig.~\ref{fig:spo_density}(b)] is well deformed and has three nodes, thereby showing its $pf$-shell nature, while at the ND minimum, it shows a $sd$-shell nature. 
  Therefore, the SD minimum appears as a result of the crossing of the $sd$-orbits and the $pf$-orbit caused by the strong deformation of the system.   


   \begin{figure}[tbp]
    \begin{center}
      \includegraphics[width=0.5\textwidth]{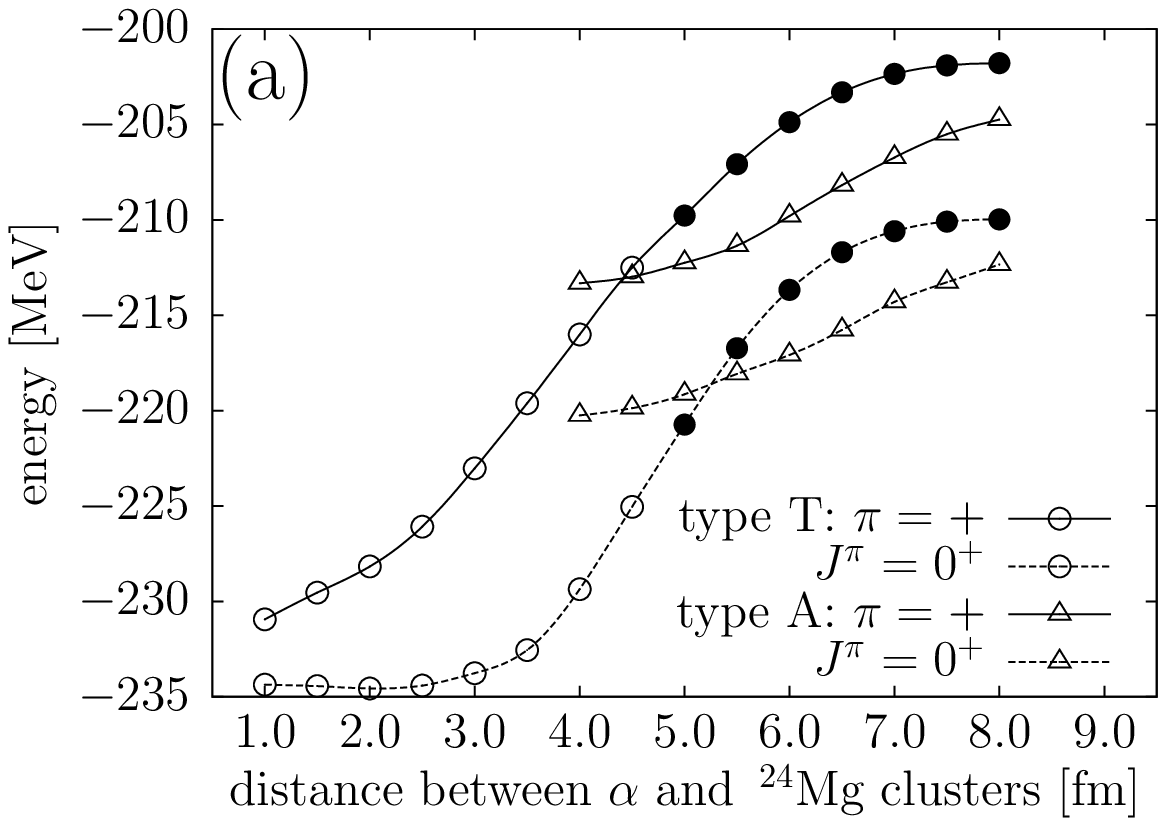}\\
      \includegraphics[width=0.5\textwidth]{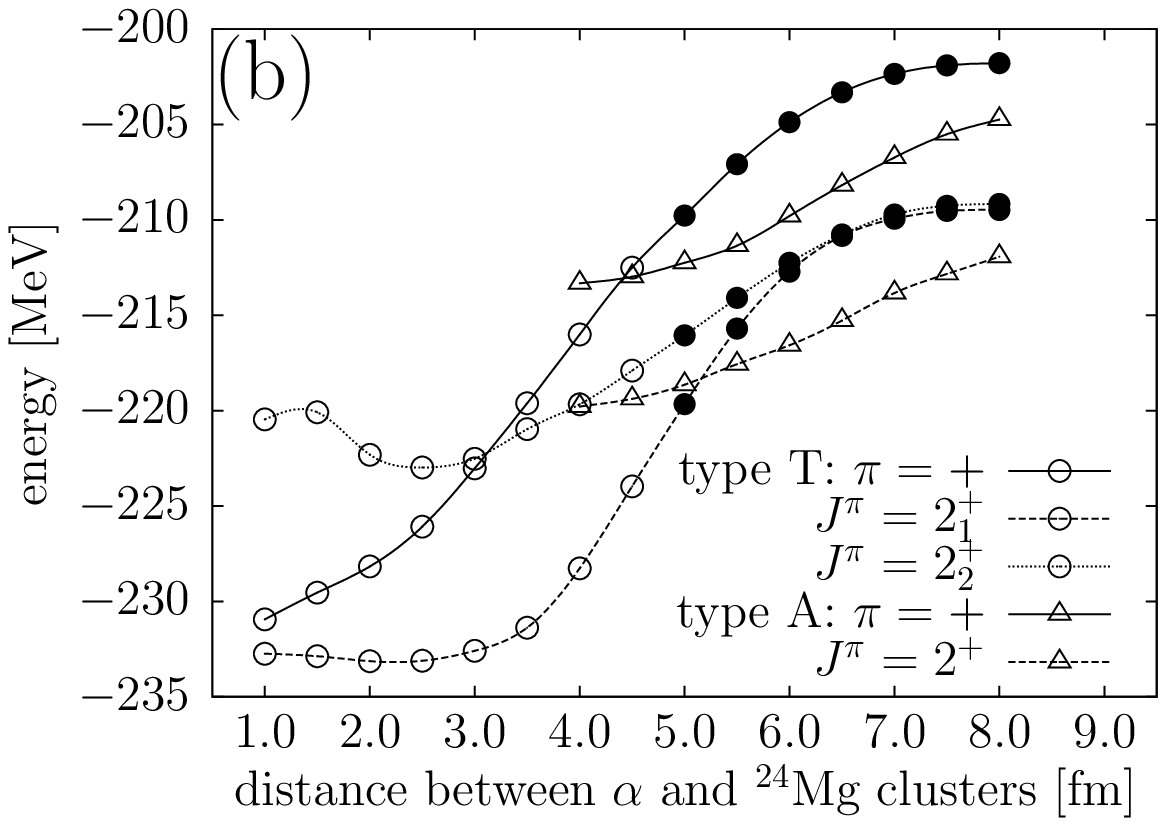}
    \end{center}
    \caption{
    $\dAMg$-energy curves projected to (a) positive-parity and $J^\pi = 0^+$ and (b) positive-parity and $J^\pi = 2^+$ states. 
    Filled circles are obtained by hand. (See text)
    }
    \label{fig:surface_A24Mg}
   \end{figure}
   
  We next discuss the results obtained by energy variation while imposing the $d$ constraint. 
  Figure~\ref{fig:surface_A24Mg} shows the energy curve obtained by the $\dAMg$ constraint. 
  The energy of the positive-parity states is shown by solid lines as functions of  the inter-cluster distance. 
  The energy curves before and after angular-momentum projection to $J^\pi = 0^+$ and $2^+$ are given by the dashed and dotted lines. 
  Two types of shapes are obtained by energy variation. One is the triaxial shape (denoted as type-T) [Fig.~\ref{fig:density}(d)] and the other is the axial symmetric shape (denoted as type-A) [Fig.~\ref{fig:density}(e)].
  The $^{24}$Mg cluster in the $\alpha$-$^{24}$Mg cluster structures deforms prolately. 
  In case of  type-A, the $\alpha$ cluster is located on the symmetric axis of the deformed $^{24}$Mg cluster, while in the case of type-T, the orientation of the deformed $^{24}$Mg cluster is transverse and the longitudinal axis of the $^{24}$Mg cluster is perpendicular to the inter-cluster direction. 
  The type-T wave functions are obtained in small $\dAMg$ region. With the increase of $\dAMg$, the structure changes from the type-T into the type-A structure. 
  Because of the triaxiality of type-T wave function, two $J^\pi = 2^+$ states are obtained by $K$-mixing as shown in Fig.~\ref{fig:surface_A24Mg}(b). 
  The energy of the second $J^\pi = 2^+$ state of type-T and that of the $J^\pi = 2^+$ state of type-A are almost the same at $\dAMg \simeq 4$ fm, but the overlap of those wave functions is quite small, and hence the type-T and -A wave functions are not mixed in the result from MCM.
  As mentioned above, in the large $\dAMg$ region, the type-A wave function is favored and the type-T wave functions are not obtained by the energy variation. 
  As shown later, the obtained type-T wave functions are found to play an important role for an $\alpha$-cluster band. 
  In order to check the behavior of the type-T structure in the large $\dAMg$ region and its effect on the band structure, we have prepared the type-T wave functions with $\dAMg=5.0-8.0$ fm by shifting by hand the $\alpha$ cluster position in the type-T wave function at $\dAMg = 4.5$ fm (filled circles in Fig.~\ref{fig:surface_A24Mg}). 


  \begin{figure}[tbp]
    \begin{center}
      \includegraphics[width=0.5\textwidth]{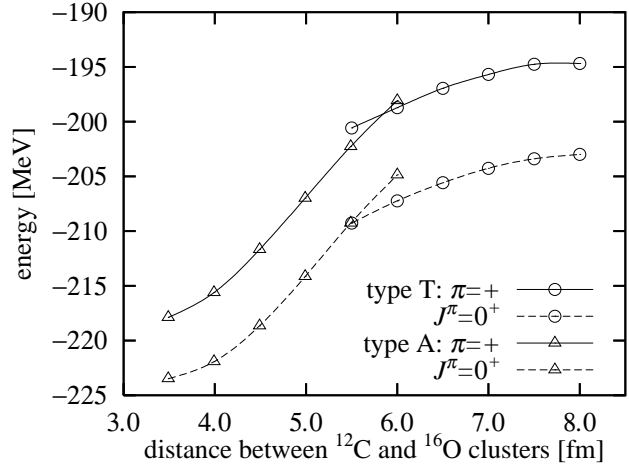}
    \end{center}
    \caption{
      $\dCO$-energy curves projected to positive-parity and $J^\pi = 0^+$ and $2^+$ states. 
    }
    \label{fig:surface_12C16O}
  \end{figure}

  Figure~\ref{fig:surface_12C16O} shows $\dCO$-energy curves projected to positive-parity and $J^\pi = 0^+$ state. 
  Two types of shapes, type-T [Fig.~\ref{fig:density}(f)] and type-A [Fig.~\ref{fig:density}(g)], are obtained due to the deformation of the $^{12}$C cluster as in the case of the $\dAMg$ constraint. 
  Namely, in the type-A wave functions, the $^{16}$O is cluster located on a symmetric axis of the oblate $^{12}$C cluster, while in the type-T wave functions, the symmetric axis of oblate $^{12}$C cluster is perpendicular to the inter-cluster direction. 
  In contrast to the $\dAMg$-energy curves, the type-T wave functions are obtained in a small $\dAMg$ region, while the type-A structure is favored in a large $\dAMg$ region. 
  The result occurs because clusters should be excited owing to the Pauli principle when the clusters overlap in a small $d$ region. 
  In order to avoid overlap, a symmetric axis of a prolate (oblate) cluster tends to be oriented perpendicular (parallel) to the inter-cluster direction in a small $d$ region. 

  In the type-A $^{12}$C-$^{16}$O wave functions with small $\dCO$, the $^{16}$O cluster is excited and forms an $\alpha$-$^{12}$C-like structure, and these wave functions are similar to the $\beta$-constrained wave functions at the ND local minimum.
  
  \section{Band structures}

  In this section, we discuss the results obtained by the MCM calculation.  \label{sec:band_structures}

  \subsection{MCM calculation and energy levels}

  We have performed the MCM calculation by using the obtained basis wave functions. 
  The adopted basis are 22 $\beta$-constrained with $\beta = 0.07$--0.48 for oblate shapes and $\beta = 0.37$--0.95 for prolate shapes, 15 type-T and nine type-A $\dAMg$-constrained wave functions with $\dAMg = 1.0$--8.0 fm and 4.0--8.0 fm, respectively, and six type-T and six type-A $\dCO$-constrained wave functions with $\dCO = 5.5$--8.0 fm and $\dCO = 3.5$--6.0 fm, respectively. 
  In the MCM calculation, $|K| \leq 2$ and $\bra{\Phirm}\hat{P}^{J^\pi}_{KK}\ket{\Phirm}/\ovlp{\Phirm}{\Phirm} > 0.005$ states are adopted as the MCM basis. 
  Other states are omitted, because they contain numerical errors in the numerical integration of the angular momentum projection. 
  The convergence of the MCM calculation is confirmed by changing the number of the basis wave functions. 
  
\begin{figure*}[tbp]
    \begin{center}
      \includegraphics[width=\textwidth]{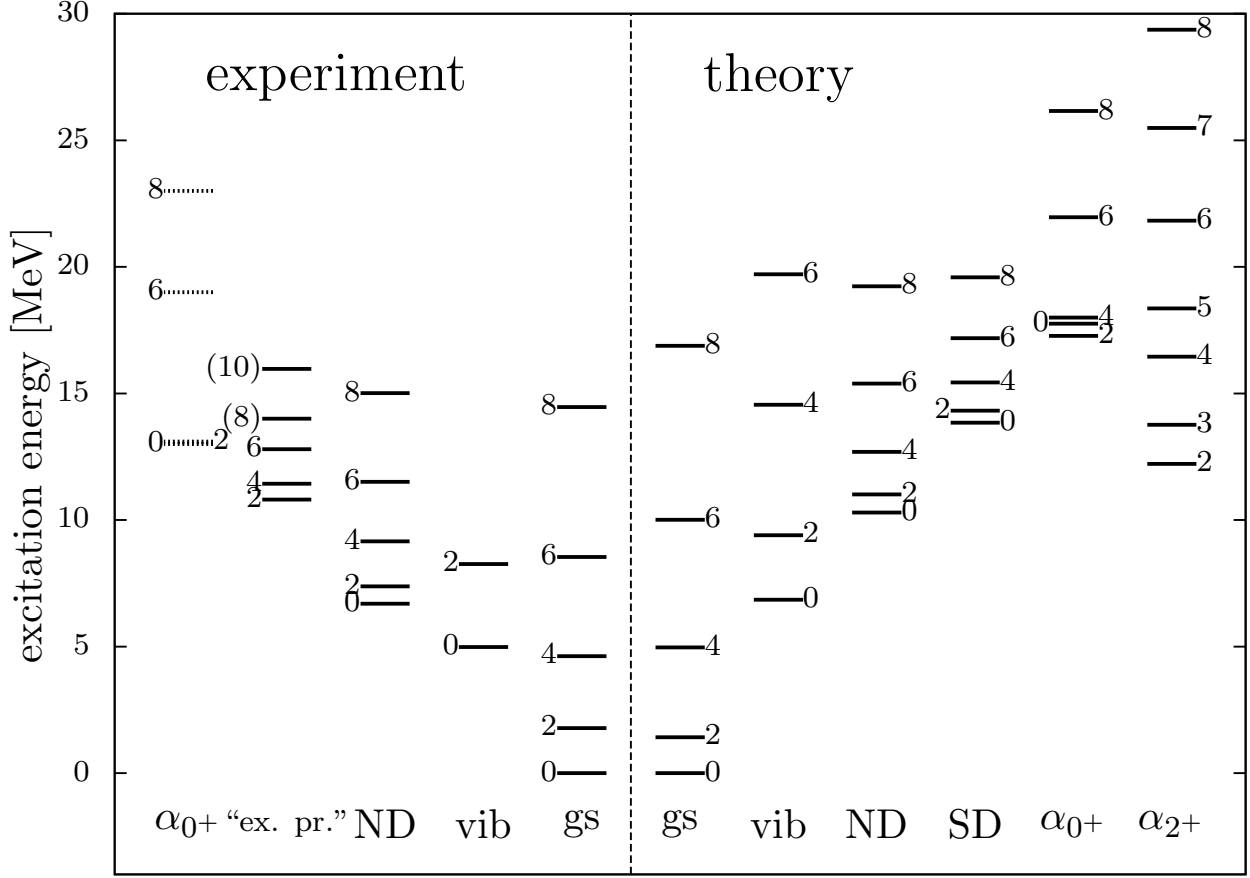}
    \end{center}
    \caption{
      Level scheme of positive-parity states in $^{28}$Si is shown. 
      Left and right panels are for experimental and theoretical values, respectively. 
      The experimental data are taken from Refs.~\onlinecite{Kubono1986461} and \onlinecite{Endt19981}. 
      The ``ex.~pr.'' indicates ``excited prolate'' band. 
    }
    \label{fig:level}
  \end{figure*}

  As shown in the energy spectra in Fig.~\ref{fig:level}, many rotational bands are constructed due to the coexistence of various structures.  An oblate ground (g) band, a $\beta$ vibration (vib) band, a ND band and a SD band are obtained. Moreover, two $\alpha$ cluster bands ($\alpha_{2^+}$ and $\alpha_{0^+}$) with $\alpha$-$^{24}$Mg cluster structure are found in high excited states. 
  As for the experimental $\alpha_{0^+}$ band, averaged values of the excitation energy are listed because those states are fragmented (dotted lines).\cite{AGG+90}   

  \subsection{Shape coexistence and $\beta$ vibration}
  \label{sec:coexistence}

  \begin{figure}[tbp]
   \begin{center}
    \includegraphics[width=0.5\textwidth]{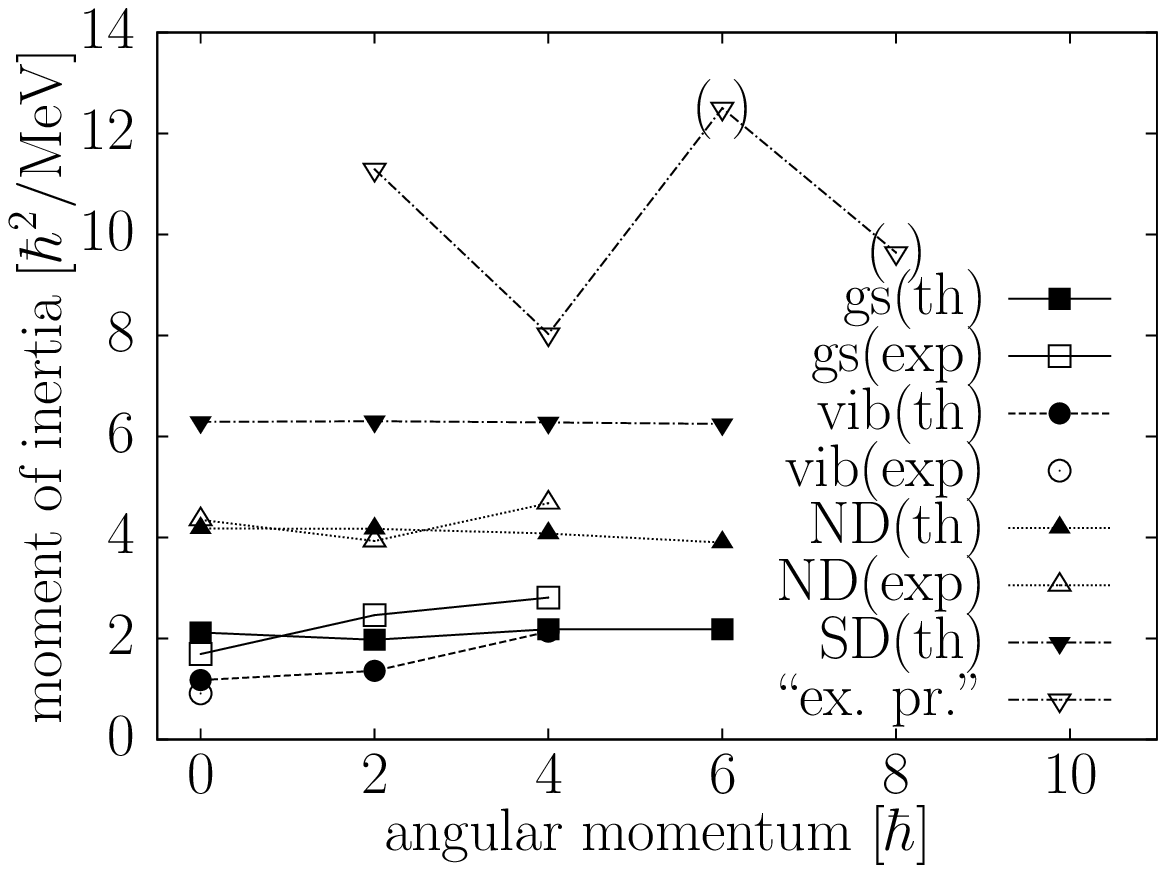}
   \end{center}
   \caption{
   Moments of inertia for the ground, $\beta$ vibration, ND and SD band. 
   Open and closed symbols indicate theoretical and experimental values, respectively. 
   The definition of moment of inertia is ${\cal J}(J) = \frac{2J + 3}{E(J+2) - E(J)} \hbar^2$, where $E(J)$ is excitation energy of the angular momentum $J$ state. 
   }
   \label{fig:moi}
  \end{figure}
  
  The excitation energies of the ground state band and $\beta$ vibration band members have good agreement with experimental data. 
  As for the ND band, the calculated excitation energies are slightly higher than the experimental ones, but the calculations reproduce well the level spacing in the ND band as shown in Fig.~\ref{fig:moi}, which shows moments of inertia as functions of angular momentum. That is, 
  both theoretical and experimental values of moments of inertia of the ND band are almost constant and approximately 4 $\hbar^2$/MeV. 
  
  \begin{figure}[tbp]
   \begin{center}
    \includegraphics[width=0.5\textwidth]{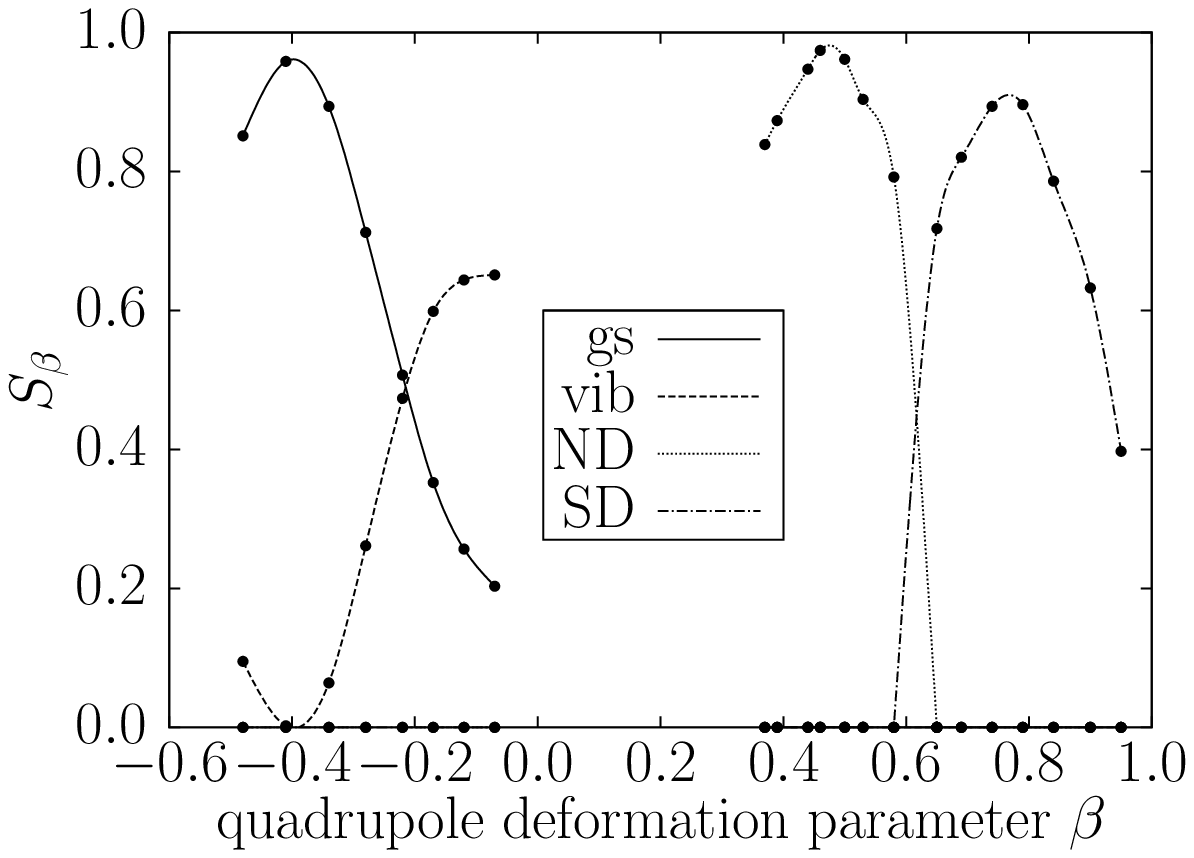}
   \end{center}
   \caption{
   Squared overlaps with $\beta$-constrained wave functions as functions of quadrupole deformation parameter $\beta$ for ground, $\beta$ vibration, ND, and SD bands are shown. 
   $\beta$ values are defined as negative values for oblate shapes. 
   }
   \label{fig:so_beta}
  \end{figure}

  In order to analyze the wave functions of the ground, $\beta$ vibration, ND, and SD bands, squared overlaps $S_{\beta = \beta_i}=|\langle\Phi^{\rm MCM}|\Phi^\beta_i\rangle|^2$ with $\beta$-constrained wave functions are used, as shown in Fig.~\ref{fig:so_beta}, where the $\beta_i$ is the value of the quadrupole deformation parameter $\beta$ for the $\ket{\Phirm^\beta_i}$. 
  The $S_{\beta = \beta_i}$ values for the band-head $0^+$ states are plotted as functions of quadrupole deformation parameter $\beta$.
  The wave functions of both the ground and $\beta$ vibration bands have large amplitudes in the oblate region, and it is found that the $J^\pi = 0_{\rm vib}^+$ state appears owing to its orthogonality to the ground state $J^\pi = 0_\mathrm{gs}^+$, which shows a $\beta$ vibration mode. 
  The ground state band amplitudes a peak at $\beta=-0.4$, which shows a rather large deformation.
  The ND and SD bands have large amplitudes at prolate regions ($\beta \simeq 0.4$ and 0.8, respectively). 
  These results suggest the shape coexistence of the oblate and prolate normal-deformations and the prolate superdeformation.

  \subsection{Superdeformed band}
  
  The present result predicts the SD band starting form $J^\pi = 0^+$ state at 13.8 MeV, 
  though the SD band has not been clearly identified. 
  There are experimental works that argue for the existence of the ``excited prolate'' band with a large moment of inertia starting from the state at around 10 MeV\cite{Kubono1986461,PhysRevC.33.1524,Kubono1981320}.
  We compare the theoretical SD band and the experimental ``excited prolate'' band.
  
  As shown in Figs.~\ref{fig:surface_beta} and \ref{fig:so_beta}, SD states are obtained by wave functions around the local minimum at $\beta \simeq 0.8$. 
  Reflecting the large deformation, the moments of inertia are large, and take a value of approximately 6 $\hbar^2$/MeV as shown in Fig.~\ref{fig:moi}. 
  However, the moments of inertia for the ``excited prolate'' band deduced from the band assignment of the experimental work are much larger than those of the theoretical SD band. 
  The present results unsupport the band assignment of ``excited prolate'' band. 
  In order to conclude the correspondence of the theoretical SD band and the ``excited prolate'' band, more experimental information, such as intra-band transitions, are required. 
  As shown in the next subsection, large strengths for the electric quadrupole transition are suggested in the present SD band.
  
  \subsection{Electric quadrupole transition strengths $B(E2)$}

  \begin{table}[tbp]
    \caption{
      Electric quadrupole transition strengths $B(E2)$ are shown. 
      Units are in Weisskopf unit $B(E2)_{{\rm W.u.}} = 5.05 e^2{\rm fm}^4$ for $^{28}$Si. 
      $J_i$ and $J_f$ are the angular momentum of initial and final states, respectively. 
      Experimental data and the results of the $^{12}\mathrm{C}+^{16}\mathrm{O}$ potential model (PM) are also listed. 
      The experimental data are taken from Ref.~\onlinecite{Endt19981}. 
    }
    \label{tab:BE2}
    \begin{center}
      \begin{tabular}{cccccc}
	\hline
	& $J_i$ & $J_f$ & experiment & present & PM\\
	\hline
	intra & $2^+_\mathrm{gs}$ & $0^+_\mathrm{gs}$  & $13.2 \pm 0.3$& 15.0\\
	& $4^+_\mathrm{gs}$ & $2^+_\mathrm{gs}$ & $13.8 \pm 1.3$ & 23.2 \\
	& $6^+_\mathrm{gs}$ & $4^+_\mathrm{gs}$ & $9.9 \pm 2.5$ & 28.6 \\
	& $8^+_\mathrm{gs}$ & $6^+_\mathrm{gs}$ & --- & 33.3 \\
	& $2^+_{{\rm vib}}$ & $0^+_{{\rm vib}}$ & $5.5 \pm 1.3$ & 8.7 \\
	& $4^+_{{\rm vib}}$ & $2^+_{{\rm vib}}$ & --- & 14.2 \\
	& $6^+_{{\rm vib}}$ & $4^+_{{\rm vib}}$ & --- & 15.2 \\
	& $2^+_{{\rm ND}}$ & $0^+_{{\rm ND}}$ & --- & 41.7 & 45.9 \\
	& $4^+_{{\rm ND}}$ & $2^+_{{\rm ND}}$ & $29 \pm 5$ & 57.4 & 63.9 \\
	& $6^+_{{\rm ND}}$ & $4^+_{{\rm ND}}$ & $> 16$ & 59.1 & 67.1 \\
	& $8^+_{{\rm ND}}$ & $6^+_{{\rm ND}}$ & --- & 56.3 & 64.9 \\
	& $2^+_{{\rm SD}}$ & $0^+_{{\rm SD}}$ & --- & 132.1 \\
	& $4^+_{{\rm SD}}$ & $2^+_{{\rm SD}}$ & --- & 188.1 \\
	& $6^+_{{\rm SD}}$ & $4^+_{{\rm SD}}$ & --- & 205.8 \\
	& $8^+_{{\rm SD}}$ & $6^+_{{\rm SD}}$ & --- & 212.6 \\
	\hline
	inter & $0^+_{{\rm vib}}$ & $2^+_\mathrm{gs}$  & $8.6 \pm 1.6$ & 6.7\\
	& $2^+_{{\rm vib}}$ & $0^+_\mathrm{gs}$  & $0.029 \pm 0.009$ & 0.3 \\
	& $2^+_{{\rm vib}}$ & $4^+_\mathrm{gs}$  & $0.8 \pm 0.3$ & 3.3 \\
	\hline
      \end{tabular}
    \end{center}
  \end{table}

  Table~\ref{tab:BE2} shows electric quadrupole transition strengths $B(E2)$ of intra- and inter-band transitions. 
  Experimental data and theoretical values of the $^{12}$C~+~$^{16}$O potential model\cite{Ohkubo2004304} are also listed. 
  
  Reflecting the large deformation of the ND and SD states, the $B(E2)$ values for their intra-band transitions are large. 
  The $^{12}$C~+~$^{16}$O potential model gives consistent results with 
the present calculation  for the intra-band transitions in the ND band.
  Compared to the experimental data, the $B(E2)$ values are overestimated. 
  Especially, $B(E2;6_\mathrm{gs}^+ \rightarrow 4_\mathrm{gs}^+)$ is much overestimated. 
  It might be because structural changes with an increase of angular momentum are not represented enough in the present framework, which uses variation before angular momentum projection. 

  \section{Cluster correlations}
  \label{sec:cluster_correlations}
  
  \begin{table}[tbp]
    \begin{center}
      \caption{
	Squared overlaps of obtained wave functions and oblate ($\beta_{\rm OB}$), prolate ND ($\beta_{\rm ND}$) and SD ($\beta_{\rm SD}$), $\alpha$-$^{24}$Mg type-T ($\alpha_{\rm T}$) and A ($\alpha_{\rm A}$), and $^{12}$C-$^{16}$O type-T (C$_{\rm T}$) and A (C$_{\rm A}$) components are shown. 
	See text about the ND and SD. 
      }
      \label{tab:so}
      \begin{tabular}{ccccccccc}
	\hline
	band	&	$J^\pi$	&	$\beta_{\rm OB}$	&	$\beta_{\rm ND}$	&	$\beta_{\rm SD}$	&	$\alpha_{\rm T}$	&	$\alpha_{\rm A}$	&	C$_{\rm T}$	&	C$_{\rm A}$	\\
	\hline
	gs	&	$0^+$	&	0.97	&		&		&	0.97	&		&		&		\\
	&	$2^+$	&	0.97	&		&		&	0.97	&		&		&		\\
	&	$4^+$	&	0.97	&		&		&	0.93	&		&		&		\\
	vib	&	$0^+$	&	0.97	&		&		&	0.89	&		&		&		\\
	&	$2^+$	&	0.95	&		&		&	0.86	&		&		&		\\
	&	$4^+$	&	0.94	&		&		&	0.53	&		&		&		\\
	ND	&	$0^+$	&		&	0.99	&		&		&		&		&	0.89	\\
	&	$2^+$	&		&	0.99	&		&		&		&		&	0.88	\\
	&	$4^+$	&		&	0.99	&		&		&		&		&	0.88	\\
	SD	&	$0^+$	&		&		&	0.94	&		&	0.88	&	0.15	&		\\
	&	$2^+$	&		&		&	0.94	&		&	0.88	&	0.16	&		\\
	&	$4^+$	&		&		&	0.94	&		&	0.88	&	0.16	&		\\
	$\alpha_{0^+}$	&	$0^+$	&	0.13	&		&		&	0.72	&		&		&		\\
	&	$2^+$	&	0.25	&		&		&	0.76	&		&		&		\\
	&	$4^+$	&	0.05	&		&		&	0.81	&		&		&		\\
	$\alpha_{2^+}$	&	$2^+$	&	0.17	&		&		&	0.98	&		&		&		\\
	&	$3^+$	&		&		&		&	1.00	&		&		&		\\
	&	$4^+$	&	0.07	&		&		&	0.99	&		&		&		\\
	\hline
      \end{tabular}
    \end{center}
  \end{table}

  Table~\ref{tab:so} shows squared overlaps of the MCM wave functions with the oblate ($\beta_{\rm OB}$), prolate ND ($\beta_{\rm ND}$) and SD ($\beta_{\rm SD}$), $\alpha$-$^{24}$Mg type-T ($\alpha_{\rm T}$) and A ($\alpha_{\rm A}$), and $^{12}$C-$^{16}$O type-T (C$_{\rm T}$) and A (C$_{\rm A}$) components.
  Here, $\beta_{\rm ND}$ and $\beta_{\rm SD}$ indicate subsets consisting of the $\beta$-constrained wave functions with $\beta = 0.37$--0.58 and $\beta = 0.65$--0.95 wave functions, respectively. 
  The definition of squared overlap is described in Sec.~\ref{subsec:spooverlap}. 
  
  The values of squared overlaps of the $\beta_{\rm OB}$ component in the ground state band are large and almost unity. 
  However, the band also contains a large amount of the $\alpha_{\rm T}$ component, which that means the degrees of freedom of the $\alpha$-$^{24}$Mg clustering are embedded in the ground state band. 
  In the $\beta$ vibration band, the values of squared overlaps of the $\beta_{\rm OB}$ component are also almost unity, similar to the ground state band. 
  The values of squared overlaps in the $\beta$ vibration band are also large but smaller than those in the ground state band, 
  which shows that the degrees of freedom of the $\alpha$-$^{24}$Mg clustering are not activated, while the $\beta$ vibration band appear by $\beta$ vibration as mentioned in Sec.~\ref{sec:coexistence}. 
  
  The states in the ND band contains a $^{12}$C-$^{16}$O cluster component in the present calculation. 
  That presence is consistent with the works of $^{12}$C~+~$^{16}$O cluster models and a $^{12}$C~+~$^{16}$O potential model, where the ND band is interpreted as the lowest band of the $^{12}$C~+~$^{16}$O cluster states.\cite{Ohkubo2004304}
  In the present work, the $^{12}$C-$^{16}$O higher nodal bands are not obtained, because the MCM basis wave functions of the type-A $^{12}$C-$^{16}$O configuration with large $\dCO$ are not sufficiently incorporated to describe the excitation of the $\CO$ relative motion. 
  
  The SD band members contain a large $\alpha_{\rm A}$ component. Since the SD states can be interpreted as the $4p4h$ states as mentioned before, it is natural that the $\alpha$-cluster correlation is enhanced in this band.  Therefore, the SD states are expected to be observed in $\alpha$-transfer reactions. 
  It should be noted that the SD state contains only the component of the type-A $\alpha$-$^{24}$Mg structure that has the very longitudinal structure. 
  This property is in contrast to the  case of the ground state band which contains only the type-T $\alpha$-$^{24}$Mg component. 
  
  The $\alpha_{0^+}$ and the $\alpha_{2^+}$ bands have the characteristic feature that they have large amounts of the type-T $\alpha$-$^{24}$Mg component, while the other components are small.  Especially the amount of the $\alpha$-$^{24}$Mg component in the $\alpha_{2^+}$ band is quite large and nearly equals unity. 
  This value indicates that this band is formed by almost pure type-T $\alpha$-$^{24}$Mg wave functions. 
  That is the reason why we call these two bands the $\alpha$ cluster bands. 
  For limitations of the present model space such as the bound state approximation, the widths or fragmentation of the $\alpha_{0^+}$ are not reproduced.

  \begin{figure}[tbp]
    \begin{center}
      \includegraphics[width=0.5\textwidth]{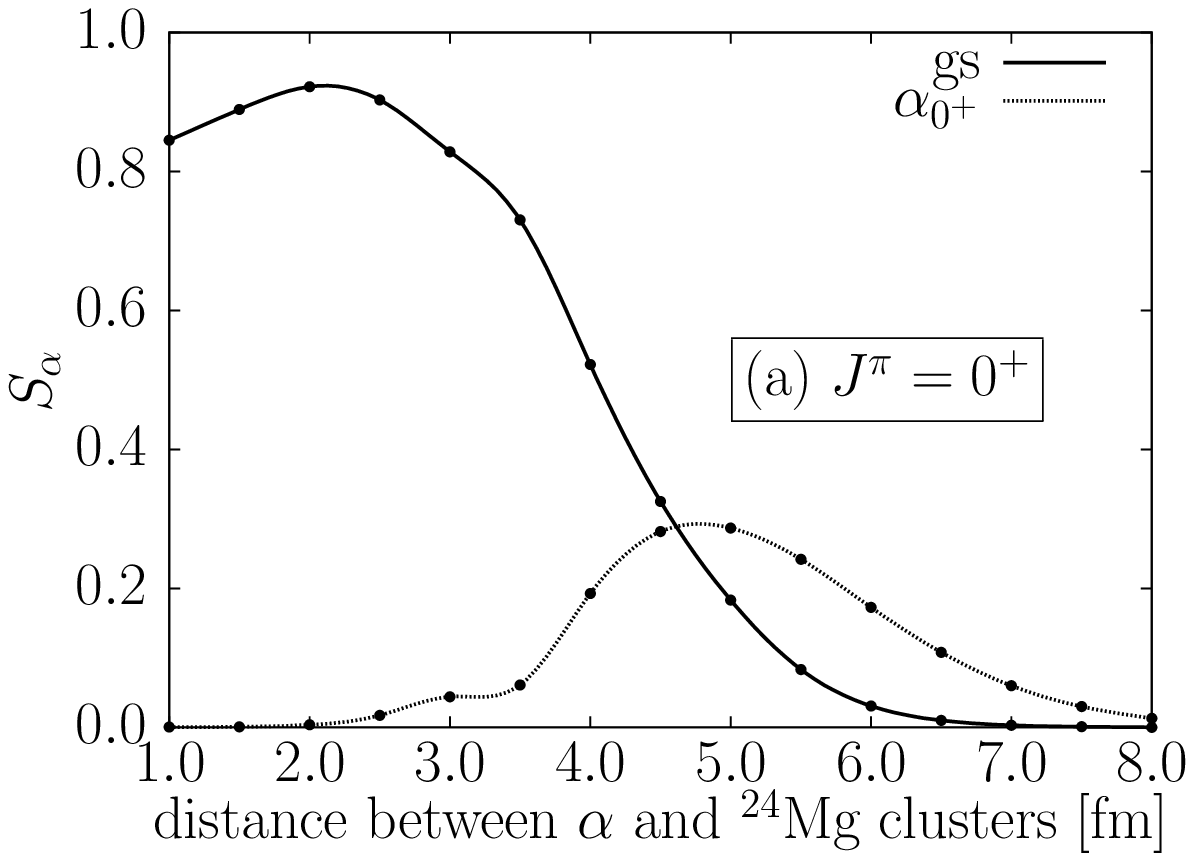}\\
      \includegraphics[width=0.5\textwidth]{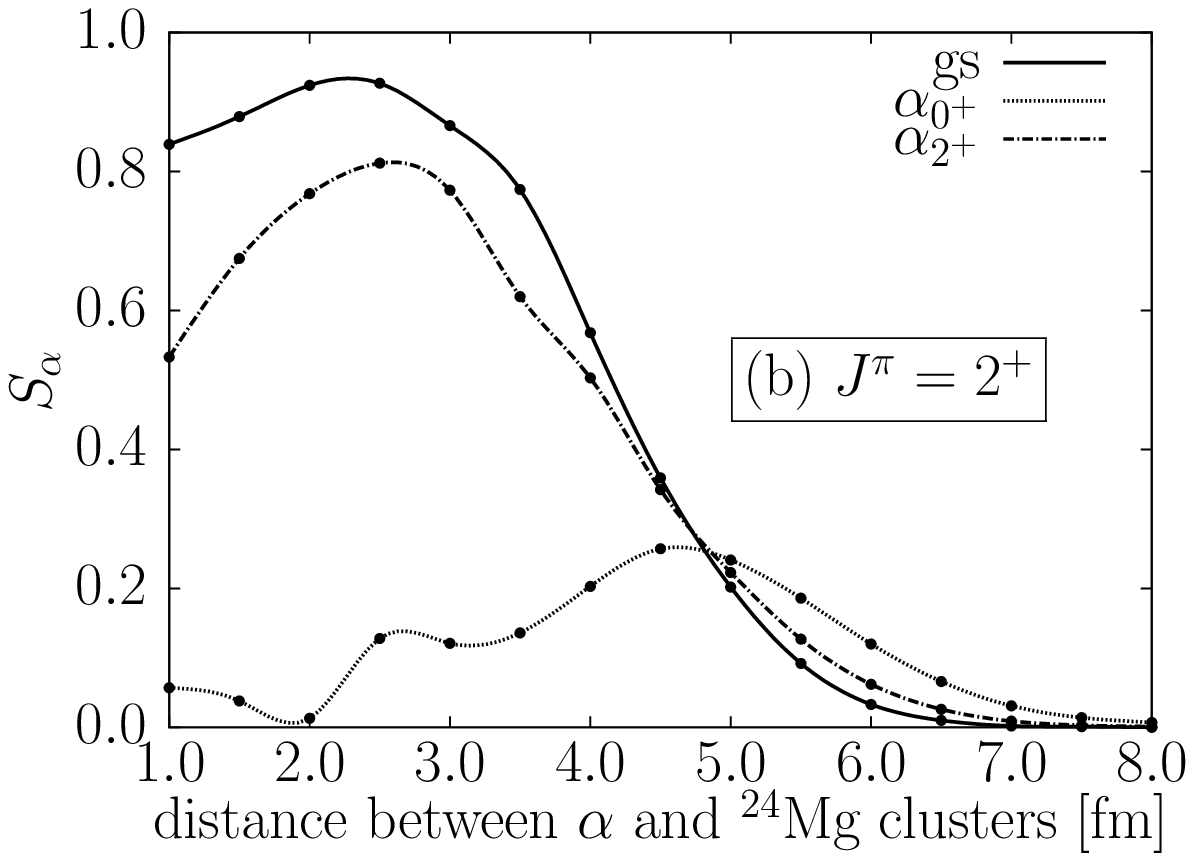}
    \end{center}
    \caption{
      $\alpha$-$^{24}$Mg cluster structure component $S_\alpha$ of the (a) $J^\pi = 0^+$ states in the gs and $\alpha_{0^+}$ bands and (b) $J^\pi = 2^+$ states in the g, $\alpha_{0^+}$ and $\alpha_{2^+}$ bands as functions of distance between $\alpha$ and $^{24}$Mg clusters. 
    }
    \label{fig:higher_nodal}
  \end{figure}
  
  Recall that the type-T $\alpha$-$^{24}$Mg  component is contained in the ground state band as well as the $\alpha_{0^+}$ and the $\alpha_{2^+}$ bands. Therefore, it is expected that these bands might be understood as cluster excited states built on the ground state band. 
  In order to analyze the cluster features of these bands, we here discuss the overlap of the states with the $\alpha$-$^{24}$Mg cluster wave functions in more detail.
  Figure~\ref{fig:higher_nodal} shows the type-T $\alpha$-$^{24}$Mg cluster structure component $S_\alpha$ of (a) the $J^\pi = 0^+$ and (b) $2^+$ states in the ground, $\alpha_{0^+}$ and $\alpha_{2^+}$ bands as functions of the distance $\dAMg$.
  As shown in the figure, the amplitudes for the $J^\pi = 0_\mathrm{gs}^+$ and $2_\mathrm{gs}^+$ in the ground state band are concentrated in the small $\dAMg$ region, while those for the $J^\pi = 0_{\alpha_{0^+}}^+$ and $2_{\alpha_{0^+}}^+$ states are compressed in the small distance region compared to the ground state band, and they have a peak at $\dAMg \sim 5$ fm. 
  This figure shows the typical $\alpha$-$^{24}$Mg higher-nodal nature of the $\alpha_{0^+}$ band built on the ground state band owing to the excitation of inter-cluster motion. 
  As mentioned before, the candidates for the members of the $\alpha$-cluster band have been observed by $^{24}$Mg($^6$Li, $d$) and $^{24}$Mg($\alpha$, $\alpha$) reactions. 
  The observed $K^\pi = 0^+$ band may correspond to the $\alpha_{0^+}$ band though the excitation energies are slightly overestimated by the present calculations.
  
  As for the $\alpha_{2^+}$ band, the $2_{\alpha_{2^+}}$  state shows a large amplitude of $S_\alpha$ in the small  $\dAMg$ region, similar to the $J^\pi = 2_\mathrm{gs}^+$ in the ground state band.
  It shows that the $\alpha_{2^+}$ band is regarded as a counter part of the ground state band owing to $K$-mixing because of the triaxial deformation of the type-T $\alpha$-$^{24}$Mg cluster structure.
  In particular, the $\alpha_{2^+}$ band has a $|K|=2$ feature of the type-T $\alpha$-$^{24}$Mg cluster structure.  
  Here, $K$ is defined with respect to the $z$-axis set to the inter-cluster direction of the $\alpha$ and $^{24}$Mg clusters, and the orientation of the prolate $^{24}$Mg cluster is perpendicular to the $z$-axis as mentioned before. 
  Therefore, the $\alpha_{2^+}$ band is interpreted as the $\alpha$-$^{24}$Mg cluster band with the cluster core excitation of $^{24}$Mg$(2^+)$.
  On the experimental side, there is no established band with $K^\pi = 2^+$. 
  In order to search for the $\alpha_{2^+}$ band, observation of unnatural parity states might be helpful.
  
  $^{28}$Si has a variety of deformed bands, and those states contain cluster components. 
  Other deformed states and cluster correlations such as a hyperdeformed states and $\alpha$-$^{20}$Ne-$\alpha$ clustering are also attractive issues\cite{Zhang199461,PhysRevC.70.034311}. 

  \section{Summary and Conclusions}
  \label{sec:summary}
  
  Positive-parity states in $^{28}$Si have been studied using the deformed-basis AMD + MCM focusing on clustering and deformation. 
The experimental energy levels in the low-energy region are reproduced well by the present calculations.
  The oblately deformed ground state band and prolately deformed excited band are reproduced, and the result shows shape coexistence. The $\beta$ vibration band also appears because the oblately deformed state is soft against quadrupole deformation. 
  A superdeformed band is suggested in the present results. 
  The SD band is described by the $(sd)^8(pf)^4$ configuration. If the suggested SD band of $^{28}$Si  is observed experimentally it should be the superdeformation of the lightest $sd$-shell nucleus. Existence of largely deformed band (``excited prolate'') has been proposed experimentally, however, we cannot assign the experimental band with the theoretical SD band because the experimental moment of inertia is not consistent with that of the calculated SD band. More experimental data such as electric or magnetic transitions are requested.

  The cluster bands, $\alpha_{0^+}$ and $\alpha_{2^+}$ bands also have been obtained. 
  These bands contain significant $\alpha$-$^{24}$Mg cluster structure components. The $\alpha_{0^+}$ band is regarded as the higher-nodal band owing to the excitation of inter-cluster motion, while the $\alpha_{2^+}$ band is interpreted as the $K^\pi = 2^+$ band due to the triaxiality of the $\alpha$-$^{24}$Mg cluster structure.

  It is found that cluster components are significantly contained in the low-lying deformed states as well as the cluster bands. Namely, the gs and SD bands contain a significant $\alpha$-$^{24}$Mg cluster structure component, and the ND band contains a $^{12}$C-$^{16}$O cluster structure component. 
  Those results are analogous to situations of other $sd$-shell nuclei such as $^{32}$S for which $^{16}$O-$^{16}$O correlations in the SD band have been suggested\cite{PhysRevC.69.051304,PhysRevC.66.021301}, and $^{40}$Ca where $\alpha$-$^{36}$Ar correlations in the ND band and $^{12}$C-$^{28}$Si correlations in the SD band have been discussed.\cite{taniguchi:044317,PhysRevC.72.064322} 

  We have shown the importance of clustering effects as well as deformation effects in low-lying states of $^{28}$Si. 
  That finding is consistent with the recent full-microscopic studies that suggested that both clustering and deformations play important roles in the wide-range $sd$-shell region.
  
  \section*{Acknowledgments}

  The authors would like to thank Dr.~M.~Takashina for fruitful discussions. 
  The numerical calculations have been carried out on SX8 at YITP, Kyoto University, SX8R at RCNP, Osaka University, and supercomputers at KEK. 
  This work has been partly supported by the Grant-in-Aid for Scientific Research from JSPS,
  and the Grant-in-Aid for the Global COE Program ``The Next Generation of Physics, Spun from Universality and Emergence'' from MEXT of Japan. 

  \bibliography{28Si_taniguchi_v2}
  
\end{document}